\documentclass[a4paper,11pt]{article}
\usepackage{amssymb}
\usepackage{amsmath}
\usepackage{caption}
\usepackage{mathrsfs}
\usepackage{commath}
\usepackage{comment}

\usepackage{url}

\usepackage[margin=3.0cm]{geometry}
\usepackage{graphicx}
\usepackage{hyperref}
\usepackage[numbers,square,sort&compress]{natbib}
\usepackage[binary-units=true]{siunitx}
\usepackage{subcaption}
\usepackage{cleveref}

\begin{document}
%------------------------------------------------------------------------------
\title{\bf Scalable computation of thermomechanical turbomachinery problems}
%------------------------------------------------------------------------------
\author{Chris N.~Richardson\thanks{BP Institute,
University of Cambridge (\url{chris@bpi.cam.ac.uk})}
\and
Nathan Sime\thanks{Department of Engineering, University of
Cambridge (now at Carnegie Institution for Science, \url{nsime@carnegiescience.edu})}
\and
Garth~N.~Wells\thanks{Department of Engineering, University of
Cambridge (\url{gnw20@cam.ac.uk})}}
\date{}

%------------------------------------------------------------------------------
\maketitle
%------------------------------------------------------------------------------
\begin{abstract}
\noindent A commonly held view in the turbomachinery community is that
finite element methods are not well-suited for very large-scale
thermomechanical simulations.  We seek to dispel this notion by
presenting performance data for a collection of realistic, large-scale
thermomechanical simulations. We describe the necessary technology to
compute problems with $O(10^7)$ to $O(10^9)$ degrees-of-freedom, and
emphasise what is required to achieve near linear computational
complexity with good parallel scaling. Performance data is presented for
turbomachinery components with up to 3.3 billion degrees-of-freedom. The
software libraries used to perform the simulations are freely available
under open source licenses. The performance demonstrated in this work
opens up the possibility of system-level thermomechanical modelling, and
lays the foundation for further research into high-performance
formulations for even larger problems and for other physical processes,
such as contact, that are important in turbomachinery analysis. \\*[2ex]
\end{abstract}

%------------------------------------------------------------------------------
\section{Introduction}
\label{sec:introduction}

There is an increasing demand for large-scale thermomechanical
simulation of turbomachinery problems. This is driven by the need for
ever tighter tolerances on deformations under thermal and mechanical
loading, and the highly integrated nature of modern designs.  The
integrated nature of advanced systems requires a move from
component-level simulation to system level simulation.  A barrier to
progress in this area, however, is the computational cost of finite
element simulation of thermomechanical problems using conventional
technology.

For a step change in capability, the important advance is towards
solvers with (near) linear complexity (time cost and memory) in problem
size, and then with good parallel scaling. A common mistake in
industrial settings, in our experience, is to focus too heavily on
parallel scaling performance and to overlook complexity. To tractably
perform very large-scale simulations the first step is the application
of methods with cost complexity that is close to linear in problem size.
Sparse direct linear solvers for three-dimensional finite element
problems have at best $O(n^{2})$ time complexity, where $n$ is the
numbers of degrees-of-freedom~\citep{george:1973,poulson:thesis}. This
is prohibitive for problems with large~$n$. The high time cost
complexity cannot be conquered by parallel implementations of direct
solvers.

The differential equations used to model thermomechanical systems are
typically elliptic or parabolic, or combinations of the two. As a
consequence, practical numerical methods must be implicit, which in turn
implies the solution of linear systems of equations. This is in contrast
with many computational fluid dynamics problems for which explicit
methods can be applied. In finite element analysis, direct sparse linear
solvers that are variants on LU decomposition, are dominant. Direct
solvers are robust, but have complexity in both time and memory that is
far from linear in problem size. In practice, with current computer
performance, advances in implementation of LU solvers, and
dimensionality and complexity effects, two-dimensional simulations are
generally tractable. However, very large three-dimensional simulations
with order $10^8$ degrees-of-freedom are made intractable by cost and
will remain so.  We explain through complexity analysis the cost
differences between two- and three-dimensional cases to elucidate why
implementation improvements alone will not make large-scale
three-dimensional analysis viable.

Extreme scale finite element simulation is advanced in some fields, such
as computational fluid dynamics, e.g.~\cite{Rasquin2014}, and
geophysics, e.g.~\cite{Rudi2015}. However, other areas are trailing,
such as thermomechanical analysis of turbomachinery. We present
computational examples of realistic thermomechanical turbomachinery
problems with up to 3.3 billion degrees-of-freedom. Our purpose is to
show that, with appropriate preconditioning, iterative linear solvers
can be effective for thermomechanical problems with complicated
geometries at extreme scale and provide a route towards whole
system/engine level analysis. Aspects of the model construction that are
critical for good and robust performance are discussed. While the cost
of the linear solver phase may be dominant in typical finite element
libraries for thermomechanical analysis, all stages of a simulation must
be considered to compute with $O(10^7)$ or more degrees-of-freedom. We
summarise the key components of an implementation that supports extreme
scale computation, and in particular the implementation used to generate
the performance data in this work.

The remainder of this work is structured as follows. The complexity of
sparse direct solvers is discussed, and this is followed by the
conditions under which iterative solvers can be applied with linear
complexity. Other performance-critical elements of large-scale
simulations are discussed briefly in the context of parallel
computation.  The description of the solver technology is followed by a
description of the physical model used in the examples, and its
numerical implementation. Performance studies are presented, and
followed by conclusions. Much of what we discuss will be familiar to
researchers in numerical linear algebra and high performance computing.
Our aim is to reach researchers and analysts working in the field of
turbomachinery to show the viability and potential of mathematically
sound methods for extreme scale computation of thermomechanical problems
using finite element methods. The computational examples that we present
are produced with freely available open-source libraries, and in
particular tools from the FEniCS Project
\citep{fenics:web,logg:2010,fenics:book,AlnaesBlechta2015a}
(\url{https://fenicsproject.org}) and PETSc
\citep{petsc_user_ref,petsc-web-page}
(\url{https://www.mcs.anl.gov/petsc/}).

%------------------------------------------------------------------------------
\section{Background: scalable approaches for large-scale problems}

Implicit finite element solvers require the solution of
\begin{equation}
  A u = b,
  \label{eqn:Ax_b}
\end{equation}
where $A$ is a $n \times n$ matrix, $u$ is the solution vector and $b$
is the right-hand side vector.  Many finite element libraries solve the
above problem using a sparse direct solver, and for large $n$ this
constitutes the most expensive part of an analysis.  A major barrier to
large-scale analysis is that sparse direct solvers have high work
complexity, especially in three-dimensions, which makes large analyses
($n > 10^{6}$) very slow and very large analyses ($n > 10^{8}$)
intractable. To enable large analyses, methods with cost close to $O(n)$
in work and storage are necessary.

In this section we describe the key elements needed to build parallel
finite element solvers for thermomechanical modelling with close to
$O(n)$ cost. First, however we summarise the performance of direct
solvers to highlight why improved implementations and parallelisaton of
direct solvers, while helpful, is a ultimately a futile endeavour in
terms of enabling very large-scale thermomechanical simulations.

%------------------------------------------------------------------------------
\subsection{Linear solvers}

%------------------------------------------------------------------------------
\subsubsection{Direct solvers: dimensionality and complexity}

Quantities of interest for characterising the cost of solving
\cref{eqn:Ax_b} are the representative length of element cells, $h$, and
the number of degrees-of-freedom in a model,~$n$. For a problem with
fixed geometric size, the number of degrees-of-freedom is proportional
to $h^{-2}$ in two dimensions and proportional to $h^{-3}$ in three
dimensions. Discretisation errors are usually characterised in terms of
$h$, and solver cost in terms of~$n$.

From \emph{a priori} error estimates, the solution error is typically
proportional to $h^{s}$, where $s \ge 1$ \citep{brenner:book}.  For a
given factor reduction in error, a three-dimensional model therefore
requires a greater relative increase in the number of degrees-of-freedom
compared to a two-dimensional problem, (by an extra power). Direct
sparse solvers have work complexity that is a power of $n$, with
the exponent depending on the spatial dimension. Solution on structured
meshes using optimal ordering requires $O(n^{3/2})$ work and $O(n \log
n)$ storage in two dimensions, and three dimensions $O(n^{2})$ work and
$O(n^{4/3})$ storage~\citep{george:1973}. Consequently, when moving from
two-dimensional to three-dimensional analysis the cost increase is
compounded two-fold; the greater increase in the number of
degrees-of-freedom for a given error reduction \emph{and} the increase
in linear solver complexity from $O(n^{3/2})$ to~$O(n^{2})$.

To make the effects of dimensionality and work complexity concrete,
consider the computation of the displacement field for a linear elastic
problem using linear Lagrange finite elements. To reduce the error in
the displacement (measured in the $L^{2}$-norm) by a factor of 10, the
cell size $h$ must be reduced by a factor of $\sqrt{10}$ ($O(h^{2})$
accuracy). For a two-dimensional problem this implies increasing the
number of degrees-of-freedom by a factor of $10$, and when accounting
for the solver complexity of $O(n^{3/2})$ the total time cost is a
factor of $10^{3/2} \approx 31.6$ greater.  In three-dimensions, the
factor increase in $n$ is 100, and after accounting for the $O(n^{2})$
solver cost the factor increase in total computational time
is~\num{10000}!

The elementary cost analysis shows that improved implementations and
parallelisaton cannot make direct solvers viable for very large problems
in three dimensions. Improved implementations can reduce the work
proportionality constant, but this will be dramatically outstripped by
the quadratic cost for large~$n$. Moreover, direct solvers are
challenging to implement in parallel. Viable solvers for large-scale
simulations must be $O(n)$, or close, in both time and memory cost.

%------------------------------------------------------------------------------
\subsubsection{Iterative solvers}

The alternative to a direct sparse solver is an iterative method. The
natural candidates are Krylov subspace methods, e.g.~the conjugate
gradient method (CG) for symmetric positive-definite operators and the
generalised minimum residual method (GMRES) for non-symmetric operators.
However, an iterative solver alone cannot noticeably improve on the work
complexity of a direct method. To illustrate, we consider as a prototype
the CG method. The core algorithmic operations are sparse matrix--vector
products and vector inner products within each iteration.  Since the
matrix $A$ is sparse, the work cost of each iteration is $O(n)$ and
storage is $O(n)$. For a CG solver to have overall work cost of $O(n)$,
it must converge in a number of iterations that is independent of~$n$.
Error analysis for the CG method shows that the number of iterations to
reach a specified error tolerance is $O(\sqrt{\kappa_{2}(A)})$, where
$\kappa_{2}(A)$ is the condition number of $A$ in the $2$-norm
\citep[Lecture~38]{trefethen:book} (this estimate is for the case when
the eigenvalues of $A$ are spread, as is the case in finite element
methods, rather than clustered).  However, the condition number of $A$
for a steady elastic problem or for the diffusion part of thermal
problem scales according to~\citep{bank:1989}
\begin{equation}
  \kappa_{2}(A) \propto n^{\frac{2}{d}},
\end{equation}
and as a result the number of CG iterations increases with mesh
refinement and the solver cost is greater than~$O(n)$.

Introducing a \emph{preconditioner} $P$, a CG solver for the problem
\begin{equation}
  P^{-1} A u = P^{-1}b,
\end{equation}
(the above is left preconditioning) will terminate in a number of
iterations that is independent of $h$ if the condition number of
$P^{-1}A$ is bounded independently of~$h$. If the preconditioner can be
applied in $O(n)$ time per iteration, the preconditioned solver will
have cost~$O(n)$. The key is to select a preconditioner $P$ that bounds
the condition number independently of $h$ (and by extension $n$).

We are aware of perceptions in the turbomachinery community that
iterative solvers for thermomechanical problems are not effective. In
our experience this comes, at least in part, from the use of algebraic
preconditioners that do not bound the condition number. For example,
incomplete Cholesky factorisation preconditioners do not, in the general
case, change the asymptotic scaling of the condition number (see,
e.g.~\cite{benzi:2002,elman:book}), which despite preconditioning,
remains~$O(h^{-2})$. The preconditioner must
account for the differential operator that generates the matrix~$A$.
Common `black-box' preconditioners, e.g.~incomplete LU, do not bound the
condition number and the number of iterations will grow with problem
size. Candidate preconditioners for the elliptic operators arising in
thermomechanical analysis are multigrid and domain decomposition
methods.

%------------------------------------------------------------------------------
\subsubsection{Multigrid preconditioned iterative solvers}

Multigrid methods \citep{briggs:book,trottenberg:book} are a natural
choice for $P$ for thermomechanical problems. While multigrid can be
used as standalone method for solving the linear system, it is more
effective for complicated engineering problems as a preconditioner for a
Krylov subspace method. Domain decomposition solvers
\citep{brandt2011revised,toselli2005domain,Mandel:2003,Farhat1991}
may also be suitable, but are generally less flexible in application
than multigrid.

Multigrid methods can be divided into two types -- geometric and
algebraic.  Geometric multigrid requires a hierarchy of finite element
meshes, preferably with a nested structure. For complicated engineering
geometries it may not be reasonable, or even possible, to produce a
hierarchy of meshes. Algebraic multigrid methods (AMG) do not require a
hierarchy of meshes, but construct a hierarchy of problems from the
`fine grid' input matrix. Common methods include classical
Ruben--St\"{u}ben AMG \citep{ruge:1987} and smoothed
aggregation~\citep{vanvek:1996}. Classical AMG tends to be suited to
scalar-valued equations, e.g.~thermal problems, and smoothed aggregation
tends to be suited to vector-valued equations, e.g.~elasticity.

A common misconception is that AMG is a \emph{blackbox} method requiring
no input or guidance from the user other than the matrix to be solved.
This is not the case, and we are aware of many cases of AMG being
mistakenly used as blackbox, and as a consequence leading to the
conclusion that it is not suitable for turbomachinery applications. For
example, the proper use of smoothed aggregation AMG requires the
`near-nullspace' of the operator to be set, which in the case of 3D
elasticity is the six rigid body modes in absence of any displacement
boundary conditions. Failure to set the full near-nullspace leads to an
increase in solver iteration count with mesh refinement. To complicate
matters, this outcome is often not observed in simple tension tests that
do not induce rotation, but the problem becomes acute when moving to
realistic analyses. This has lead to erroneous conclusions that a solver
is suitable for simple problems but not for realistic engineering
systems. It has been our experience that analysts rarely configure AMG
properly, and in particular smoothed aggregation AMG. This has
contributed to the pessimistic view of iterative solvers in the
turbomachinery community.

%------------------------------------------------------------------------------
\subsection{Cell quality}
\label{sec:cell_quality}

Orthodoxy in solid mechanics finite element analysis is to manage down
the number of degrees-of-freedom to reach a tractable cost. For
complicated geometries, this will typically compromise on cell quality
in parts of a domain. While the computational cost of direct solvers
does not depend on cell quality, the performance of preconditioned
iterative solvers is highly dependent on cell quality. A small number of
poor quality cells can dramatically slow, or prevent, convergence of an
iterative method. We refer to the discussion in~\citep{shewchuk2002good}
for an overview of finite element cell quality measures. Additionally
the work in~\citep{Klingner:2007:ATM} which highlights the challenges
in generating meshes composed of good quality cells.

The successful application of iterative solvers requires high quality
meshes, and analysts generating meshes need to refocus their efforts
away from managing the cell count and towards the generation of high
quality meshes. A contributing factor in poor experiences with iterative
solvers is the use of meshes with poor quality cells. We have observed
this to be particularly the case when attempting to benchmark against
established codes with direct solvers, where, by necessity, the cell
count is managed down to permit execution of the direct solver.

An important practical question is `how good' must a mesh be to be
acceptable. Such guidance is necessary for analysts generating meshes
for use with iterative solvers, and is a topic of ongoing investigation.

%------------------------------------------------------------------------------
\subsection{Other library components}

While the solution of a linear system may be the dominant cost when
using a direct solver, the solution of very large-scale problems
requires careful consideration and design of all stages in the solution
pipeline. Addressing the linear solver in isolation is not sufficient.
We summarise briefly other performance critical phases in a simulation
as implemented in the open-source library
DOLFIN~\citep{AlnaesBlechta2015a,fenics:web}, which is used for the
example simulations.

%------------------------------------------------------------------------------
\subsubsection{Input/output}

Input/output (IO) is often overlooked when considering extreme scale
simulations. It must be possible to read and write files in parallel,
and in a way that is memory scalable. IO must not take place on a single
process or compute node.  Some commonly used input formats for finite
element analysis are fundamentally unsuited to parallel processing.
ASCII formats do not lend themselves to efficient~IO. The examples we
present use XDMF~\citep{XDMF} with HDF5~\citep{HDF5} storage. The HDF5
files are read and written in parallel, using MPI-IO (transparently to
the user) and exploiting parallel file systems~\citep{richardson:2013}.

%------------------------------------------------------------------------------
\subsubsection{Mesh data structures}

Efficient data structures for storing the unstructured meshes associated
with turbomachinery problems are essential for scalable solution.
Examples of their implementation are shown
in~\citep{KnepleyKarpeev09,logg2009efficient}. Additionally, it is
critical that these mesh data structures are fully distributed in
parallel. Storing a mesh on one process, or a copy of the mesh on each
process, prohibits the solution of very large problems. The examples we
present use a fully distributed
mesh~\citep{richardson:2013,AlnaesBlechta2015a}, with the mesh
partitioning across processes computed using the parallel graph
partitioner PT-SCOTCH~\cite{SCOTCH}.

%------------------------------------------------------------------------------
\subsubsection{Assembly}

Scalable matrix assembly builds on a fully distributed mesh data
structure. It is essential that the matrix/vector assembly process is
fully distributed with each process responsible for assembly over its
portion of the mesh. A number of libraries exist that provide
distributed matrix and vector data structures, and these typically also
support distributed construction in which each process adds its
contribution to the distributed matrix/vector, with a synchronisation
phase to communicate any nonlocal contributions. The examples we present
use PETSc~\cite{petsc_user_ref} for distributed matrices and vectors.

%------------------------------------------------------------------------------
\section{Thermomechanical model and numerical formulation}
\label{sec:problem_definition}

We consider a thermoelastic problem on a body $\Omega \subset
\mathbb{R}^{3}$ with boundary $\Gamma := \partial \Omega$ and outward
unit normal vector $\boldsymbol{n}$ on the boundary. The boundary is
partitioned into $\Gamma_D$ and $\Gamma_N$ such that $\Gamma_D \cup
\Gamma_N = \Gamma$ and $\Gamma_D \cap \Gamma_N = \emptyset$. The time
interval of interest is $Q = (0, t_M]$.

For the thermomehcanical model, in the absence of inertia effects, the
mechanical part of the solution is governed by:
\begin{alignat}{2}
  - \nabla \cdot \sigma & = \boldsymbol{f}  && \quad \text{in} \ \Omega,
  \label{eq:elasticity_bvp}
  \\
  \sigma \cdot \boldsymbol{n}
      & = p_\text{bc} \boldsymbol{n} && \quad \text{on} \ \Gamma^{\boldsymbol{u}}_N,
  \\
  \boldsymbol{u}
    & = \boldsymbol{u}_\text{bc} && \quad \text{on} \ \Gamma^{\boldsymbol{u}}_{D},
  \label{eq:u-dbc}
\end{alignat}
where $\sigma$ is the stress tensor, $\boldsymbol{f}$ is a prescribed
body force, $p_{\rm bc}$ is a prescribed boundary pressure,
$\boldsymbol{u}$ is the displacement field and $\boldsymbol{u}_{\rm bc}$
is a prescribed displacement. The stress is given by
\begin{equation}
  \sigma = \mathcal{C} : (\epsilon - \epsilon_{T}),
\end{equation}
where $\mathcal{C}$ is the elastic stiffness tensor, $\epsilon =
\nabla^{s} \boldsymbol{u}$ is the symmetric gradient of the
displacement, and $\epsilon_{T}$ is the thermal strain. The thermal
strain is given by
\begin{equation}
  \epsilon_{T} = \alpha_L \del{T - T_{\rm ref}} I,
\end{equation}
where $\alpha_L$ is the thermal expansion coefficient, $T$ is the
temperature, $T_{{\rm ref}}$ is a fixed reference temperature and $I$ is
the identity tensor.

The temperature field $T$ is governed by
\begin{alignat}{2}
  \rho c_{v} \dpd{T}{t}
  - \nabla \cdot \kappa \nabla T & = 0  && \quad \text{in} \ \Omega \times Q,
  \\
  \kappa \nabla T \cdot \boldsymbol{n}
      & = \beta \del{T - T_{\text{bc}}}
      && \quad \text{on} \ \Gamma^{T}_{N} \times \overline{Q},
  \\
  T(x, 0)   & = T_{0}(x) && \quad \text{in} \ \overline{\Omega},
\end{alignat}
where $\rho$ is the mass density, $c_{v}$ is the specific heat, $\kappa$
is the thermal conductivity, $\beta$ is a heat transfer coefficient,
$T_{\rm bc}$ is the prescribed exterior temperature and $T_{0}$ is the
initial temperature field.

We will consider problems where $\mathcal{C}$, $\alpha_L$, $c_{v}$ and
$\kappa$ are temperature dependent. The temperature dependencies lead to
a nonlinear problem.

%------------------------------------------------------------------------------
\subsection{Fully-discrete formulation}

Applying the $\theta$-method for time stepping, the finite element
formulation reads: given $\boldsymbol{u}_{m}$ and $T_{m}$ at time
$t_{m}$, and $\boldsymbol{f}_{m + 1}$, $p_{\text{bc}, m+1}$ and
$T_{\text{bc}, m+1}$ at time $t_{m+1}$, find $\boldsymbol{u}_{m+1} \in
\left[V_{h}\right]^3$ and $T_{m+1} \in V_{h}$ such that
\begin{multline}
  F_{u}(u_{m+1}; v) := \int_\Omega \sigma(\boldsymbol{u}_{m+1}, T_{m+1}) : \nabla \boldsymbol{v} \dif x
  - \int_\Omega \boldsymbol{f}_{m+1} \cdot \boldsymbol{v} \dif x
  \\
  - \int_{\Gamma^{\boldsymbol{u}}_{N}} p_{\text{bc}, m+1} \boldsymbol{n}
    \cdot \boldsymbol{v} \dif s = 0 \quad \forall \boldsymbol{v} \in \left[V_{h}\right]^3,
\label{eq:wf_elastic}
\end{multline}
\begin{multline}
F_{T}(T_{m+1}; s) := \int_{\Omega} \rho c_{v, \theta} \frac{T_{m + 1} -
  T_{m}}{\Delta t_{m}} s \dif x
+ \int_\Omega \kappa_{\theta} \nabla T_{\theta} \cdot \nabla s \dif x
\\
- \int_{\Gamma^{T}_{N}} \beta_{\theta} \left(T_{\theta} - T_{\text{bc}, \theta}\right) s \dif s
= 0 \quad \forall s \in V_{h},
\label{eq:wf_thermal}
\end{multline}
where $w_{\theta} = (1 - \theta) w_{m} + \theta w_{m+1}$ for a
field~$w$, $t_{m+1} = t_{m} + \Delta t_m$, and $V_{h}$ is a finite
element space. In all examples we use conforming Lagrange finite element
spaces on meshes with tetrahedral cells.

%------------------------------------------------------------------------------
\subsection{Solution strategy}

The coupled system in \cref{eq:wf_elastic,eq:wf_thermal} is nonlinear
due to the dependence of various coefficients on the temperature.  The
coupling between the mechanical and thermal equations is one-way; the
mechanical equation, \cref{eq:wf_elastic}, depends on temperature but
the thermal equation, \cref{eq:wf_thermal}, does not depend on the
displacement field. Therefore, at each time step we first solve the
thermal problem ($F_{T} = 0$) using Newton's method, followed by
solution of the linear mechanical problem using the most recently
computed temperature field. The system solver strategy can be considered
a block nonlinear Gauss--Seidel iterative process.

A full linearisation of the thermal problem in \cref{eq:wf_thermal}
leads to a non-symmetric matrix operator. The degree to which symmetry
is broken depends on the magnitude of~$\partial \kappa / \partial T$.
Despite the loss of symmetry in the nonlinear thermal problem, we still
observe that the CG method performs well for the problems we consider.
There may be cases where it is necessary to switch to a Krylov solver
that is not restricted to symmetric operators, such as BiCGSTAB or
GMRES.

%------------------------------------------------------------------------------
\subsection{Adaptive time step selection}

The time step $\Delta t_{m}$ is selected to limit the maximum absolute
temperature change anywhere in the domain over a time step. The next
time step to be used is given by
\begin{equation}
  \Delta t_{m+1}
  = \frac{ \varepsilon \Delta T^{\text{max}}}{ \max_{\Omega} \left| T_{m+1} - T_m \right|} \Delta t_m,
\label{eq:dt_spec}
\end{equation}
where $\Delta T^{\text{max}}$ is the maximum permitted change in
temperature between time steps and $\varepsilon \in (0, 1]$ is a
parameter that sets the target maximum change in temperature $\Delta
T^{\text{target}} = \varepsilon \Delta T^{\text{max}}$. If the
temperature exceeds the maximum allowed the step is repeated,
with the time step halved.

%------------------------------------------------------------------------------
\section{Model problems}
\label{sec:model_problems}

\begin{figure}
  \centering
  \begin{subfigure}{0.49\textwidth}
    \centering
    \includegraphics[height=0.175\textheight]{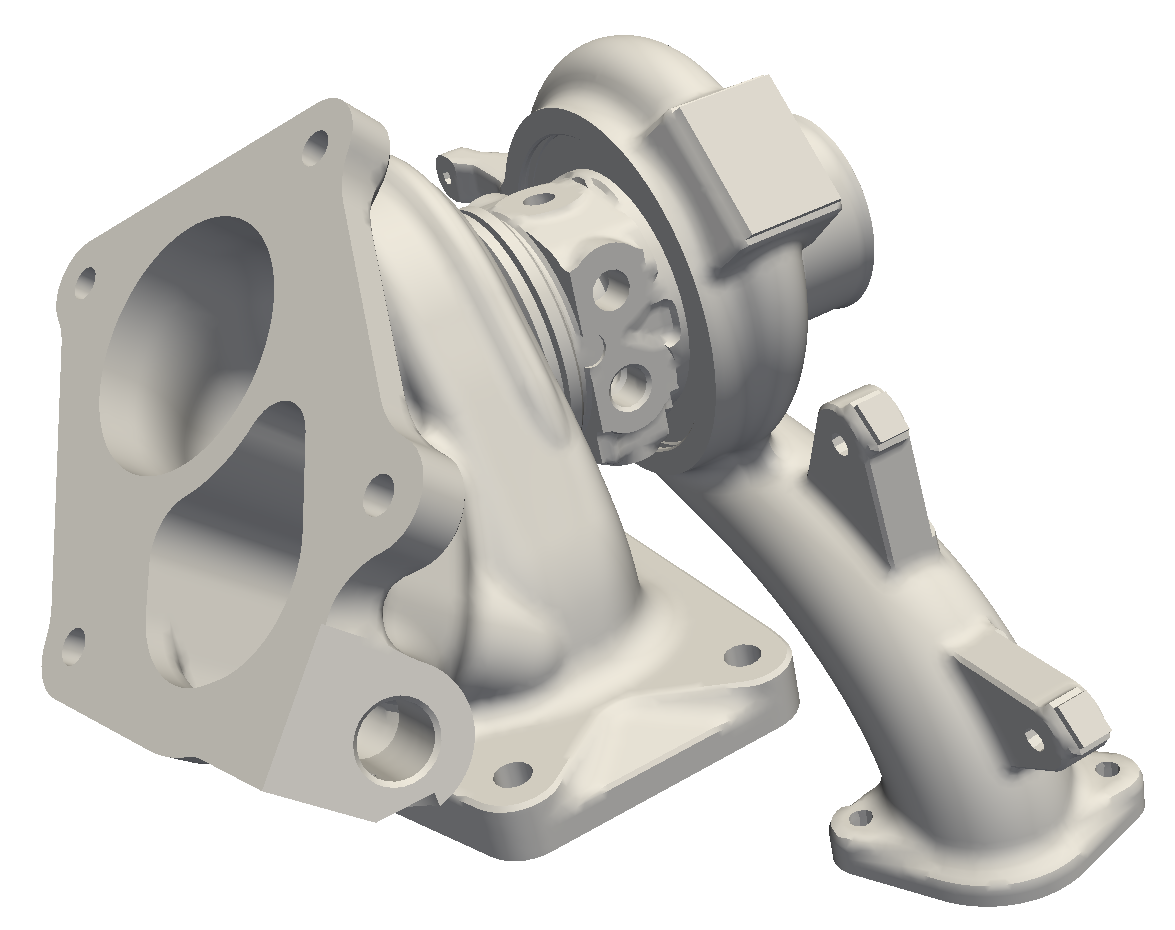}
    \caption{Exterior view.}
  \end{subfigure}
  \begin{subfigure}{0.49\textwidth}
    \centering
    \includegraphics[height=0.175\textheight]{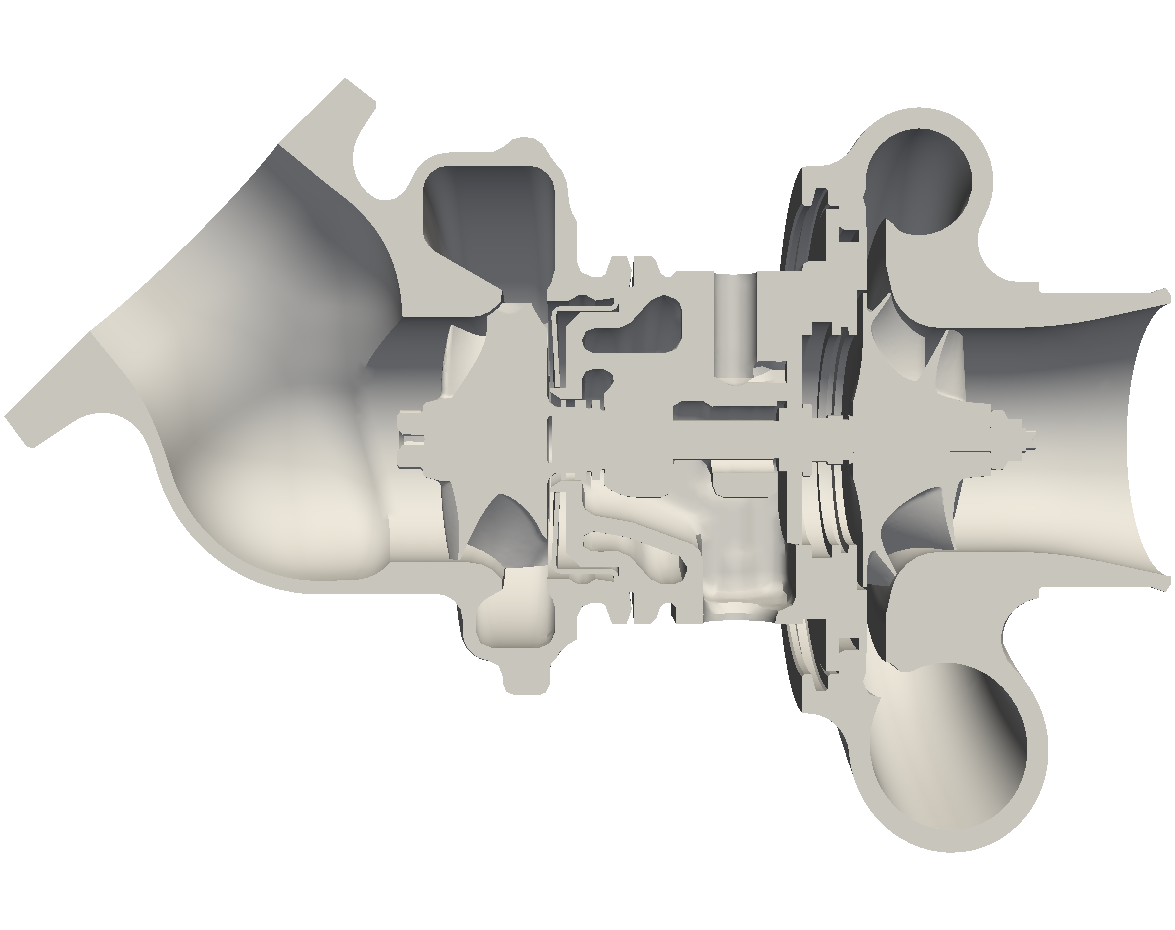}
    \caption{Cut-away view.}
  \end{subfigure}
  \begin{subfigure}{0.49\textwidth}
    \centering
    \includegraphics[height=0.175\textheight]{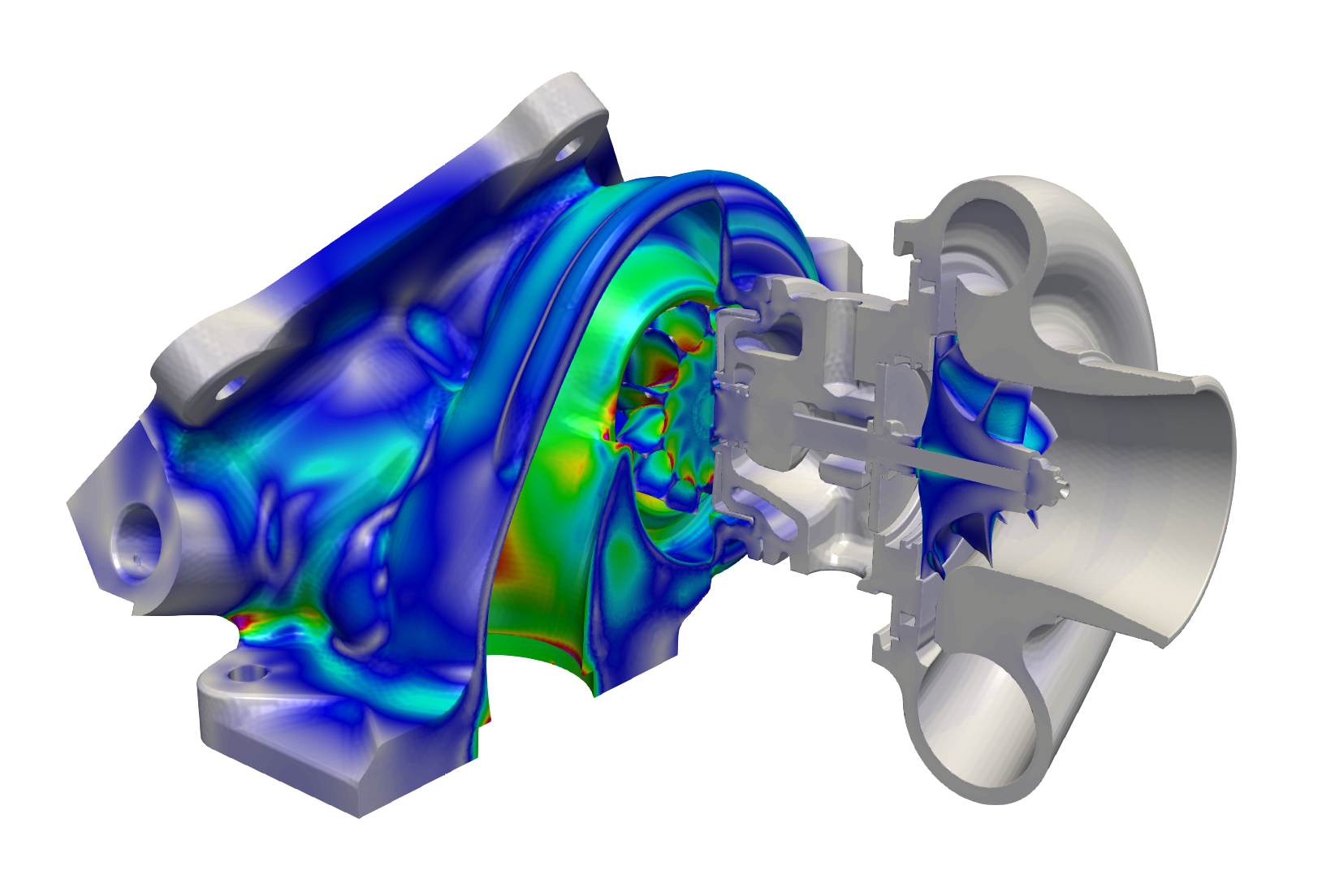}
    \caption{Look-through view (colours shows higher stress regions).}
  \end{subfigure}%
  \caption{Turbocharger geometry.}
  \label{fig:tc_domain}
\end{figure}

\begin{figure}
  \centering
  \begin{subfigure}{0.49\textwidth}
    \centering
    \includegraphics[height=0.175\textheight]{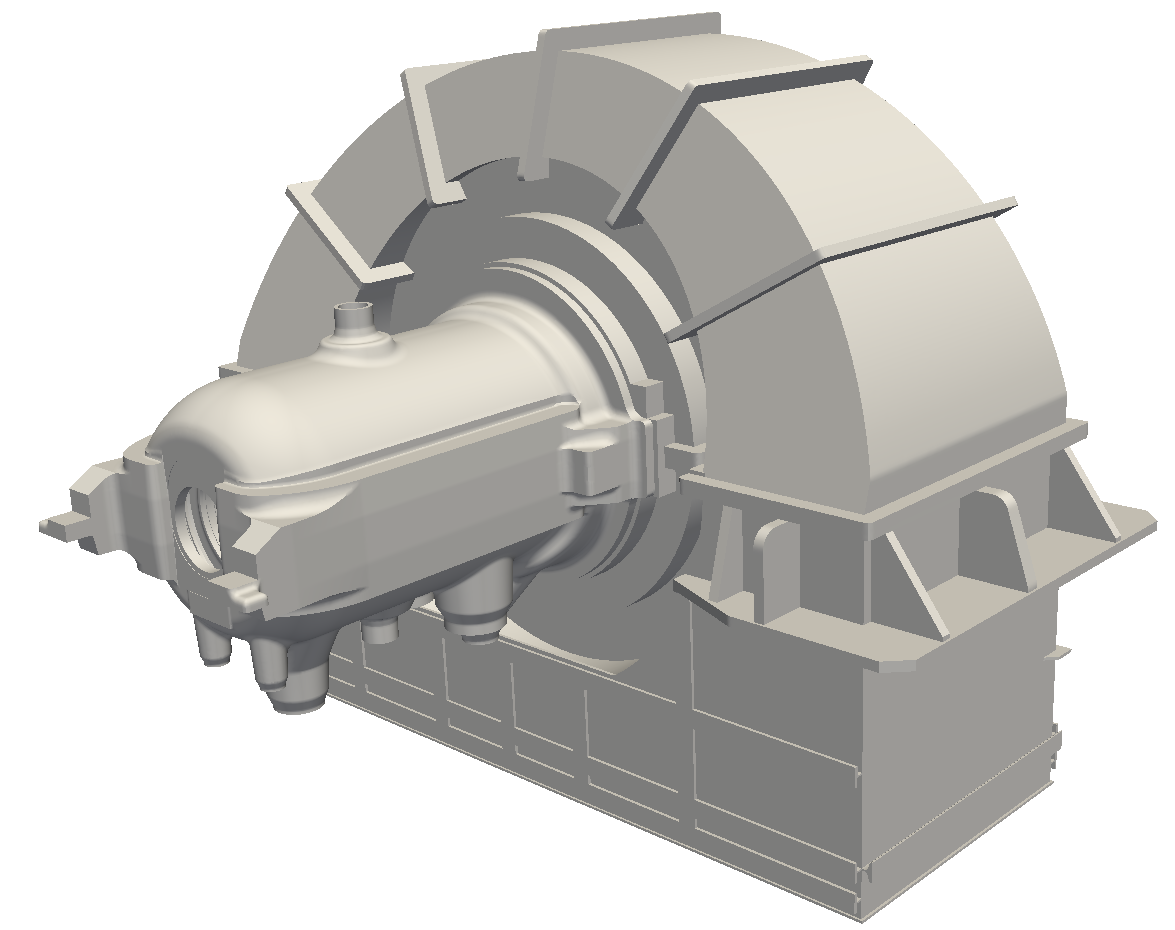}
    \caption{Exterior view.}
  \end{subfigure}%
  \begin{subfigure}{0.49\textwidth}
    \centering
    \includegraphics[height=0.175\textheight]{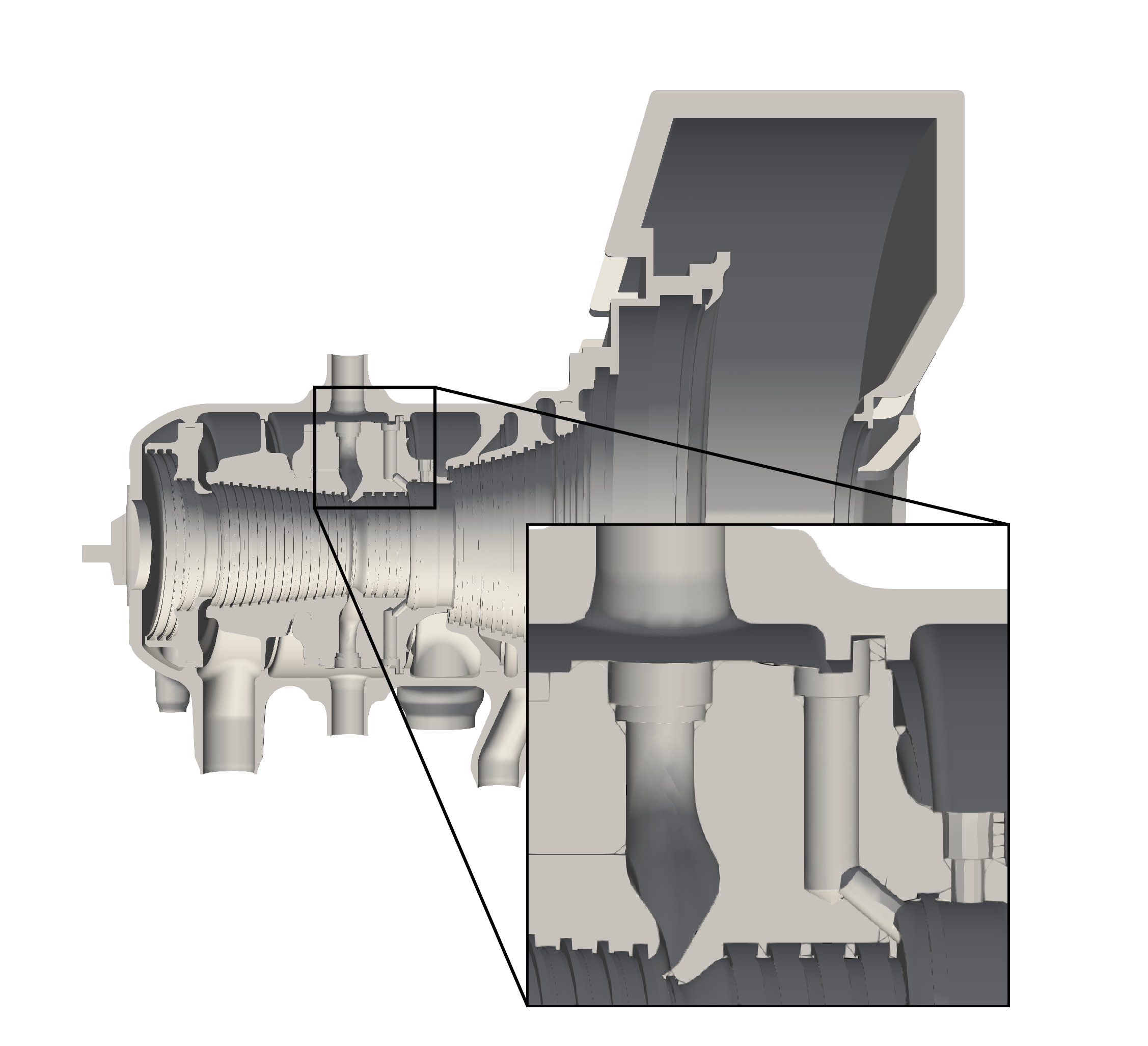}
    \caption{Cut-away view.}
  \end{subfigure}
  \begin{subfigure}{0.49\textwidth}
    \centering
    \includegraphics[height=0.175\textheight]{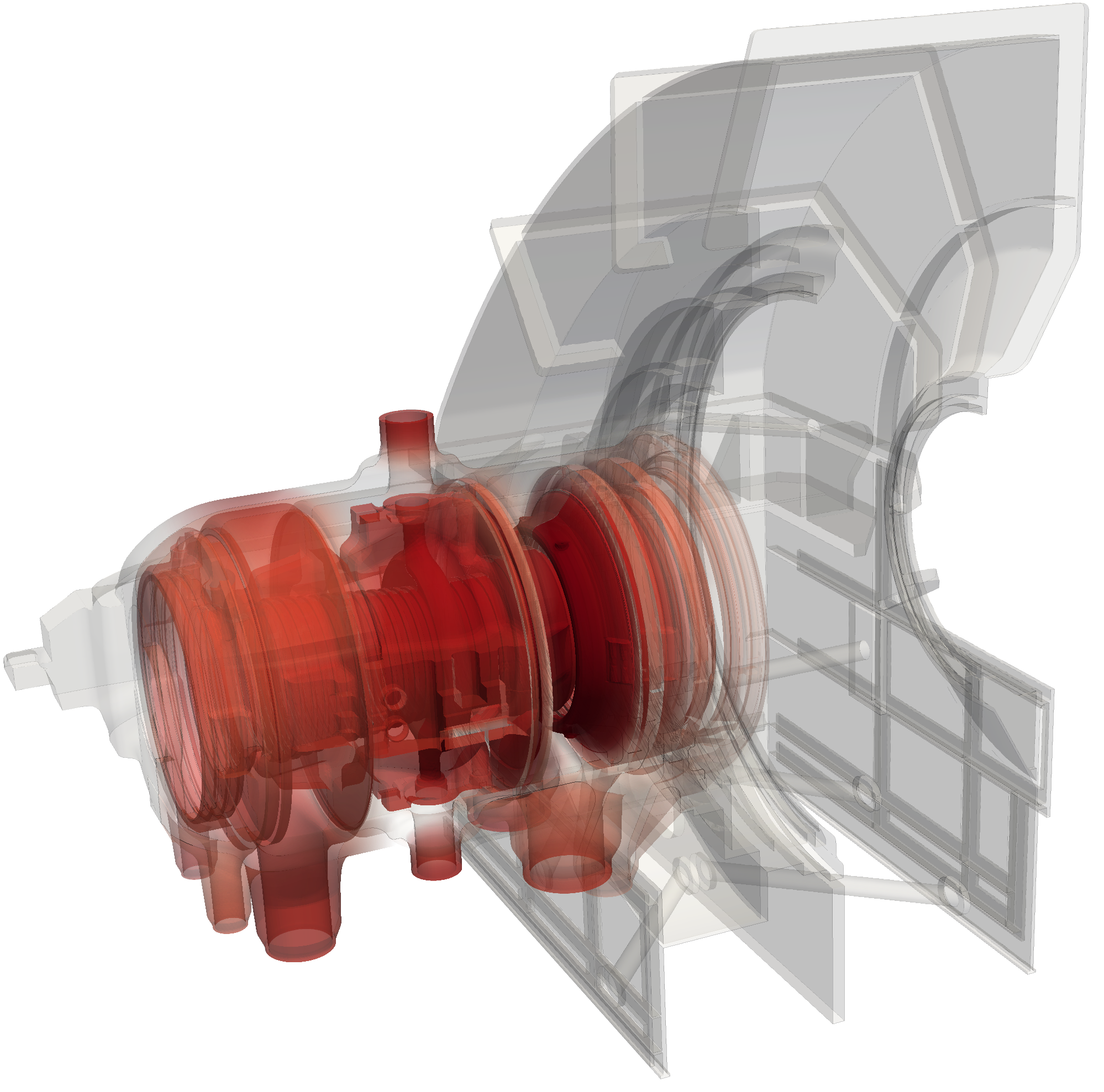}
    \caption{Look-through view (colours shows region of high temperature).}
  \end{subfigure}
  \caption{Steam turbine geometry.}
  \label{fig:st_domain}
\end{figure}

Two model problems are considered; a turbocharger (shown in
\cref{fig:tc_domain}) and a steam turbine casing (shown in
\cref{fig:st_domain}). Both problems are composed of multiple materials
with temperature-dependent thermal and elastic parameters. For unsteady
cases the boundary conditions and the applied body force are
time-dependent.

%------------------------------------------------------------------------------
\subsection{Meshes}

\begin{figure}
  \centering
  \begin{subfigure}{0.4\textwidth}
    \centering
    \includegraphics[width=0.9\linewidth]{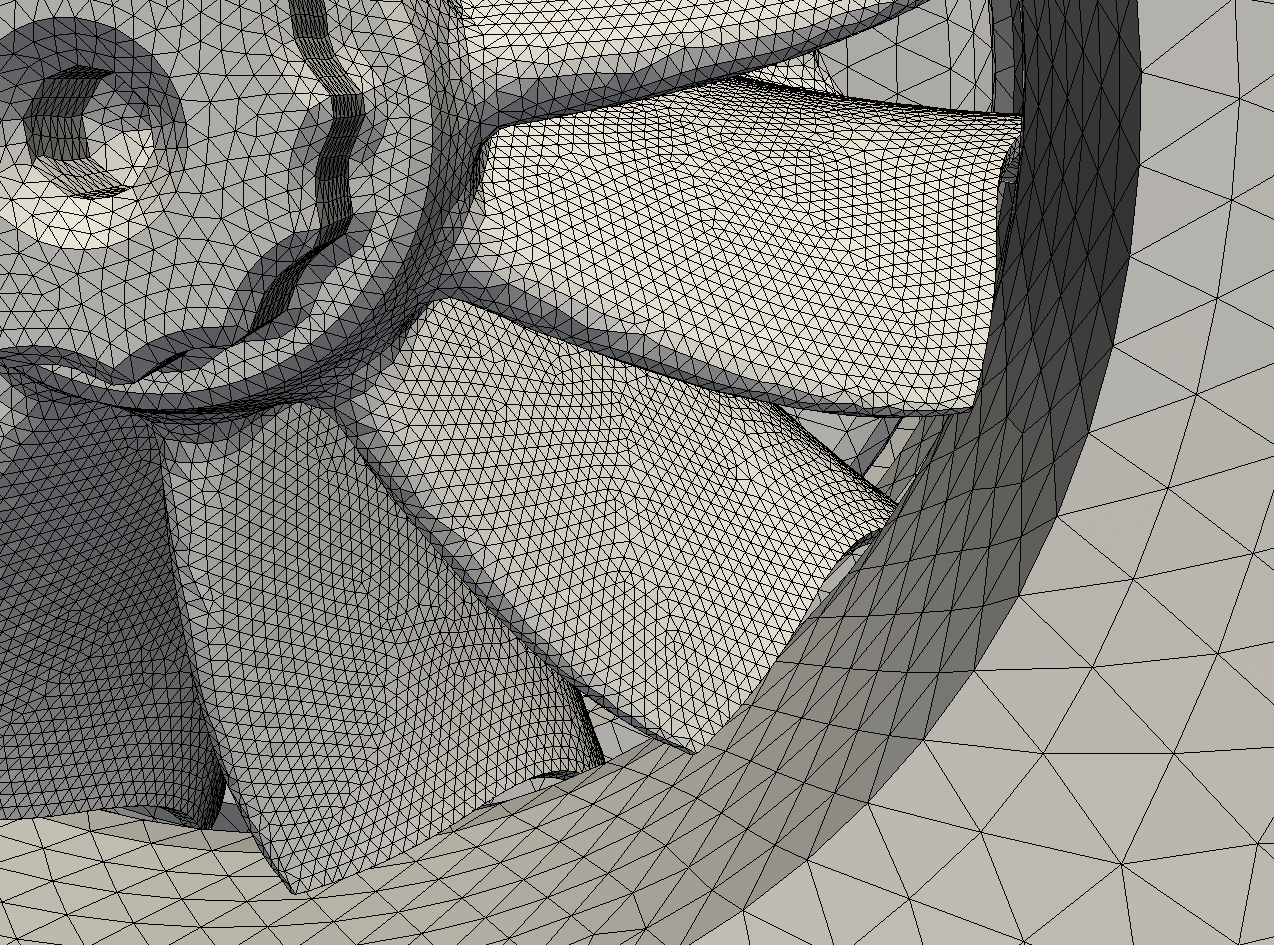}
    \caption{Original mesh.}
    \label{fig:tc_ref_0}
  \end{subfigure}%
  \begin{subfigure}{0.4\textwidth}
    \centering
    \includegraphics[width=0.9\linewidth]{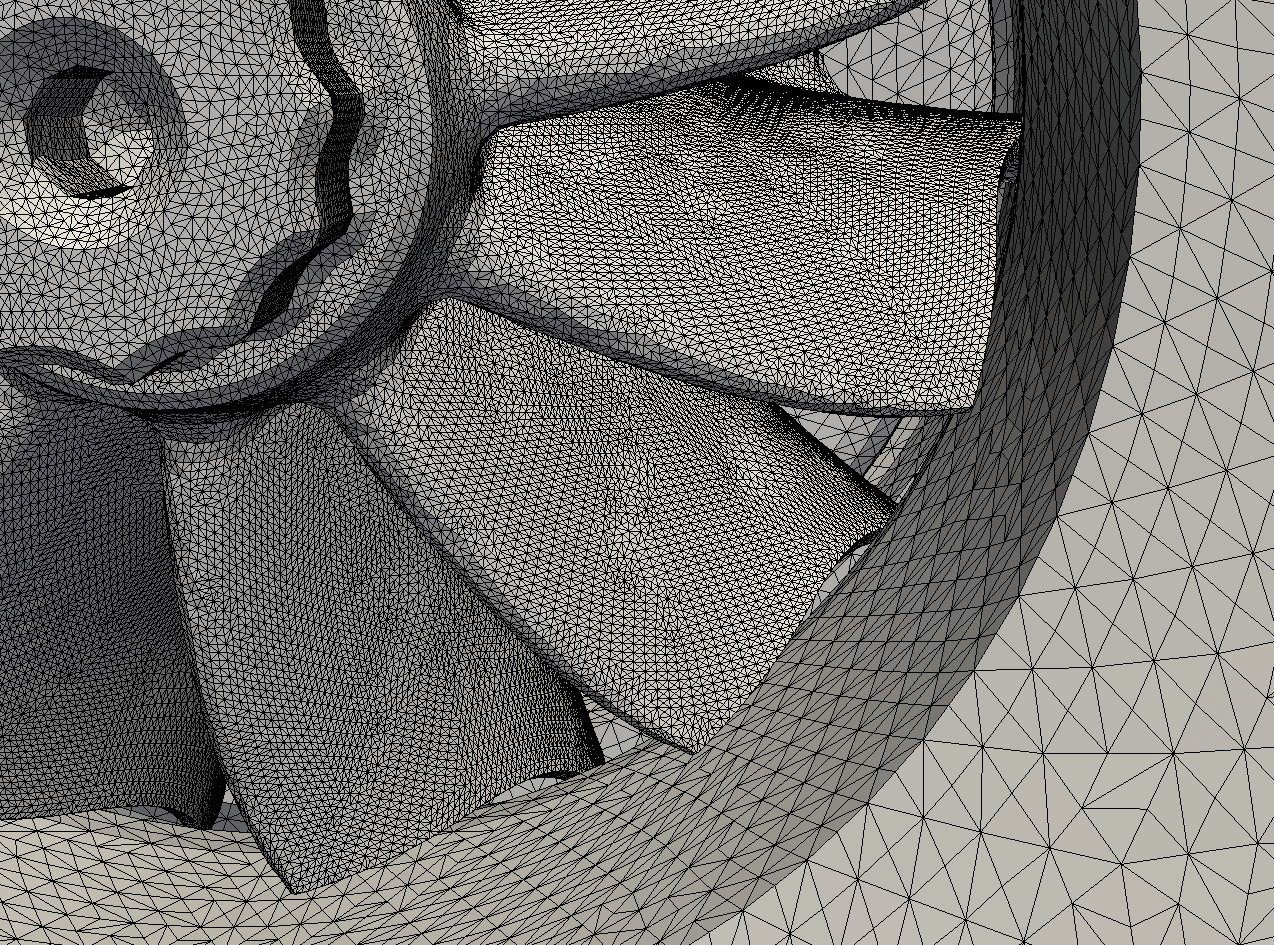}
    \caption{One level of refinement.}
    \label{fig:tc_ref_1}
  \end{subfigure}
  \begin{subfigure}{0.4\textwidth}
    \centering
    \includegraphics[width=0.9\linewidth]{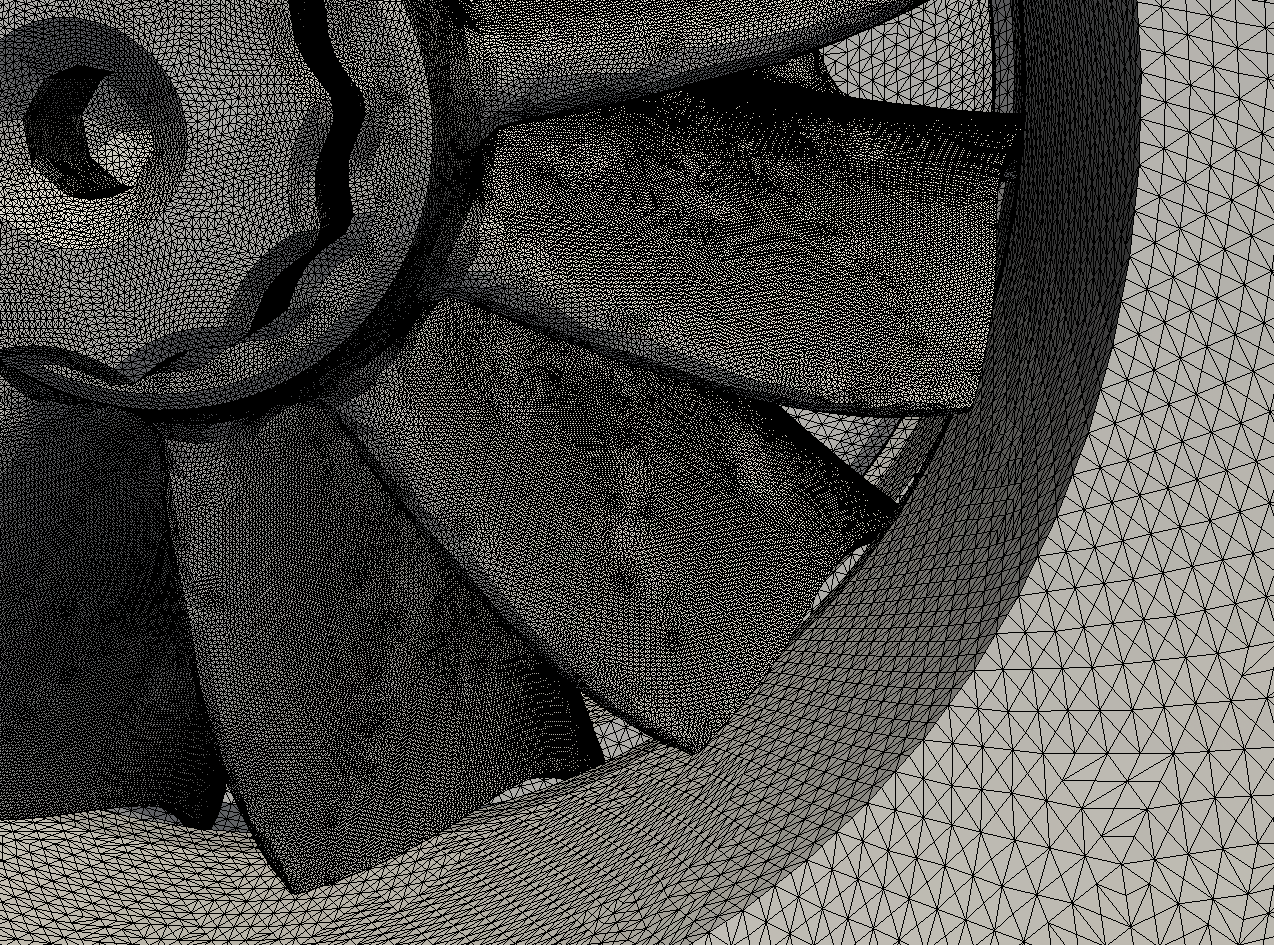}
    \caption{Two levels of refinement.}
    \label{fig:tc_ref_2}
  \end{subfigure}
  \caption{Turbocharger mesh~`A' detail around the turbine for the
    reference, once refined and twice refined mesh.}
  \label{fig:tc_mesh_detail}
\end{figure}

The meshes of the two problems are composed of tetrahedral cells. Two
`reference' meshes are considered for the turbocharger:
\begin{description}
  \item[Turbocharger mesh~`A'] \num{1603438}~cells and
  \num{375352}~vertices
  \item[Turbocharger mesh~`B'] \num{45377344}~cells and
  \num{9302038}~vertices
\end{description}
and one reference mesh for the steam turbine:
\begin{description}
\item[Steam turbine mesh] \num{25402220}~cells and
  \num{4918704}~vertices
\end{description}
Meshes with higher cell counts are constructed by applying parallel
uniform mesh refinement to the reference meshes~\citep{richardson:2013}.
Each level of uniform refinement increases the cell count by a factor of
eight. Views of a series of meshes generated in this manner are shown in
\cref{fig:tc_mesh_detail}. The algorithm used for uniform refinement is
based on the work of~\citep{plaza2000,plaza2003}. The algorithm cost is
linear with the number of cells to be refined. The difference in
tetrahedron cell quality (shape measures) between refinement levels is
small, as shown in \cref{fig:tc_mesh_detail}.

\begin{figure}
  \centering
  \begin{subfigure}{0.49\textwidth}
    \centering
    \includegraphics[width=0.9\linewidth]{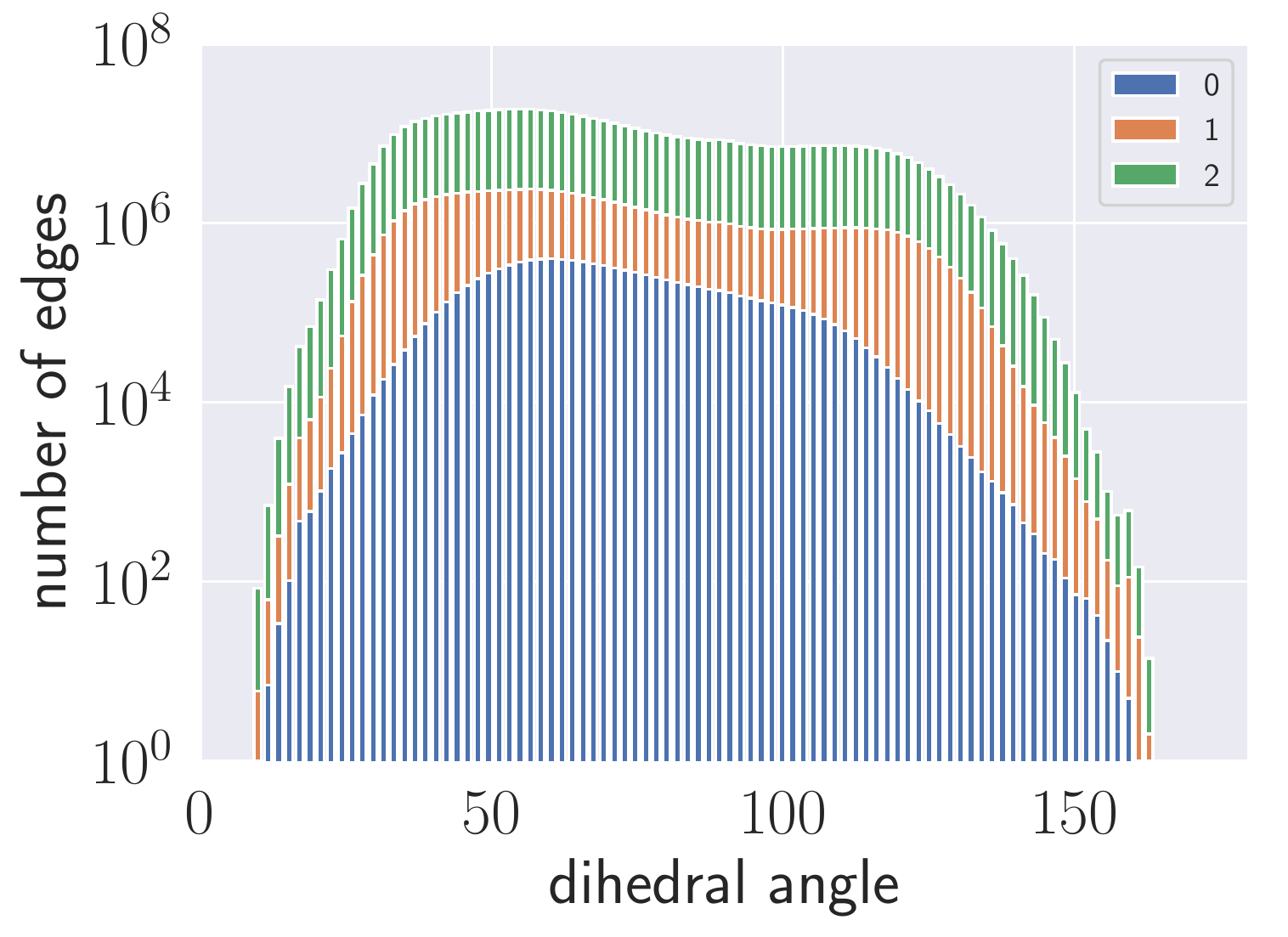}
    \caption{Turbocharger mesh~`A', \num{1603438}~cells and \num{375352}~vertices.}
    \label{subfig:tc_mesh_quality_A}
  \end{subfigure}
  \begin{subfigure}{0.49\textwidth}
    \centering
    \includegraphics[width=0.9\linewidth]{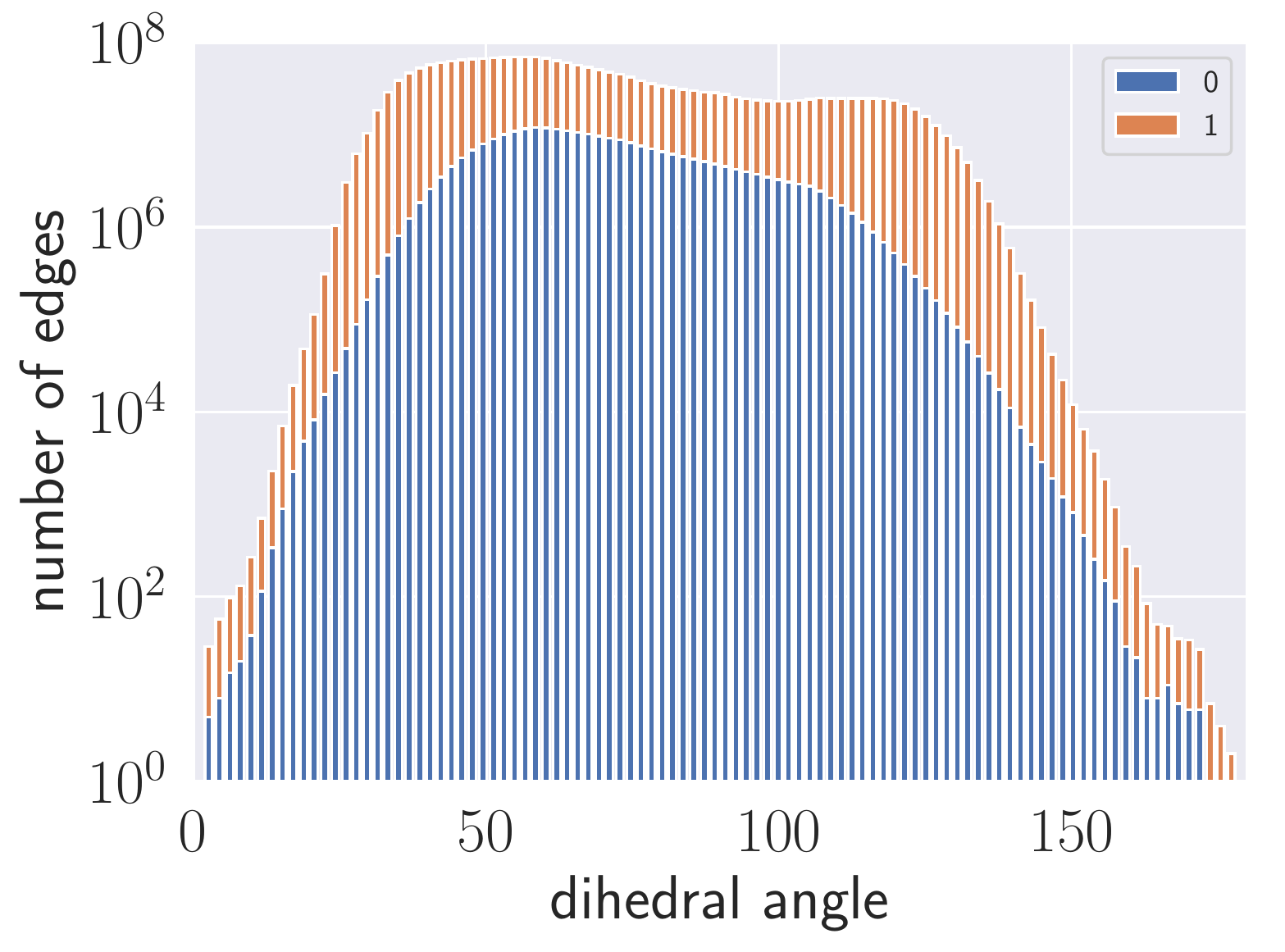}
    \caption{Turbocharger mesh~`B', \num{45377344}~cells and \num{9302038}~vertices.}
    \label{subfig:tc_mesh_quality_B}
  \end{subfigure}
  \begin{subfigure}{0.49\textwidth}
    \centering
    \includegraphics[width=0.9\linewidth]{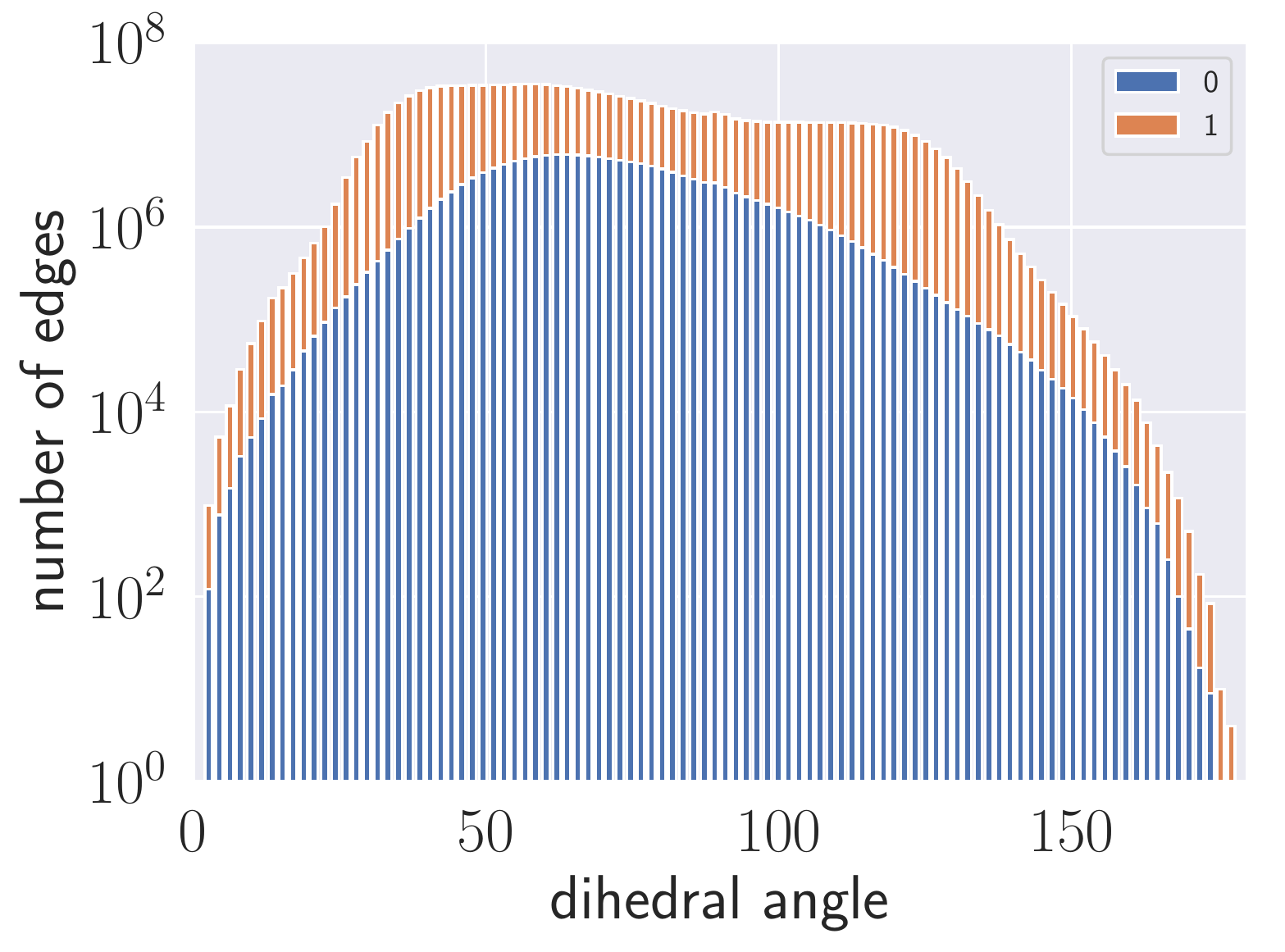}
    \caption{Steam turbine mesh.}
    \label{subfig:st_mesh_quality}
  \end{subfigure}
  \caption{Histograms of the dihedral angles for the tubomachinery meshes.
  The legend indicates the number of levels of refinement. Level~\num{0}
  corresponds to the original reference mesh.}
  \label{fig:mesh_quality}
\end{figure}

Cell quality is important for performance and robustness of the
iterative solvers. Histograms of cell quality, measured by the dihedral
angle, are shown in \cref{fig:mesh_quality} for the reference meshes. We
consider turbocharger mesh~`A' to be `high' quality (see
\cref{subfig:tc_mesh_quality_A}).  Turbocharger mesh~`B' and the steam
turbine mesh are of lower quality in this metric. We observe in our
numerical experiments that best results in terms of iteration count are
obtained using mesh~`A'. The steam turbine geometry is characterised by
a number of thin regions, and thin regions make the avoidance of poorly
shaped cells more difficult. The quantitative understanding of the
relationship between cell quality metrics and solver performance for
practical applications is an area of ongoing research.

%------------------------------------------------------------------------------
\subsection{Material parameters and boundary conditions}

\begin{figure}
  \center
  \includegraphics[width=0.45\linewidth]{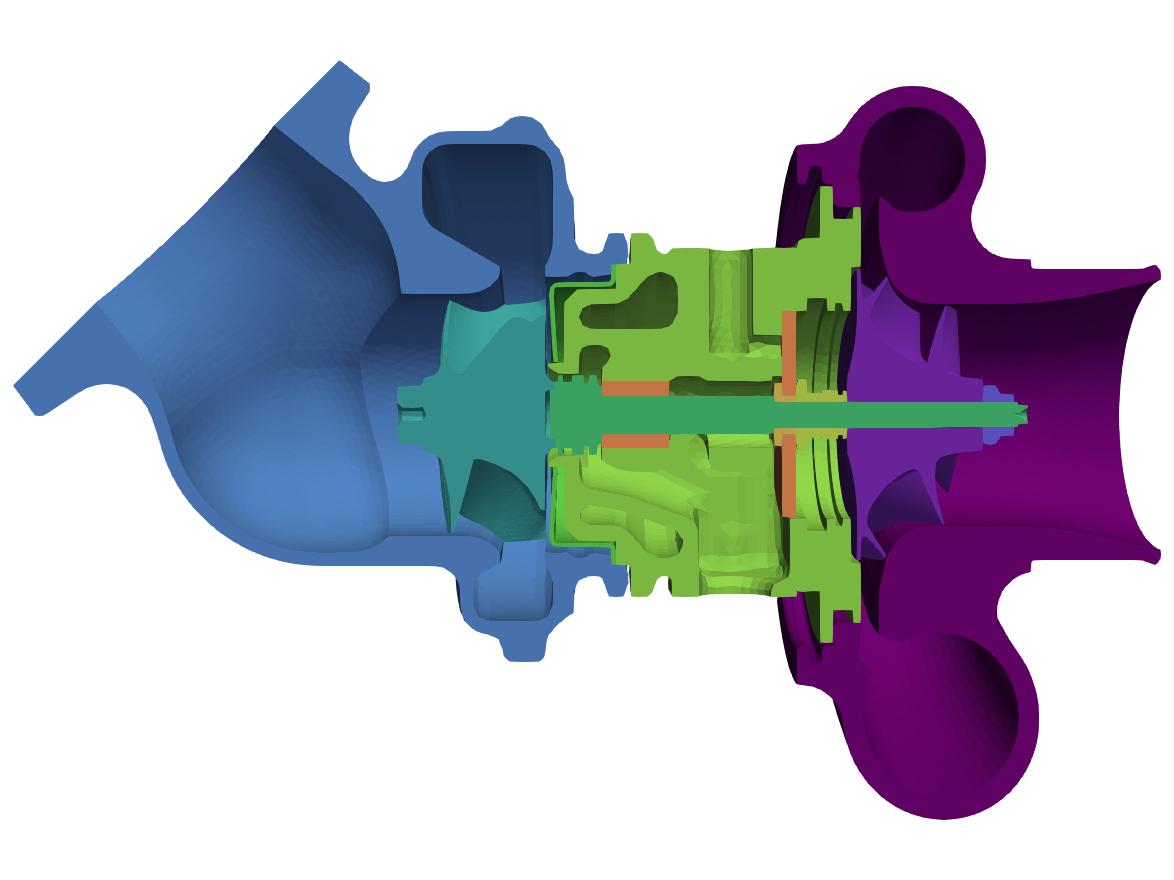}
  \caption{Illustration of different material regions. This
    problem has 12 different materials.}
  \label{fig:tc_subdomains}
\end{figure}

\begin{figure}
  \center
  \includegraphics[width=0.45\linewidth]{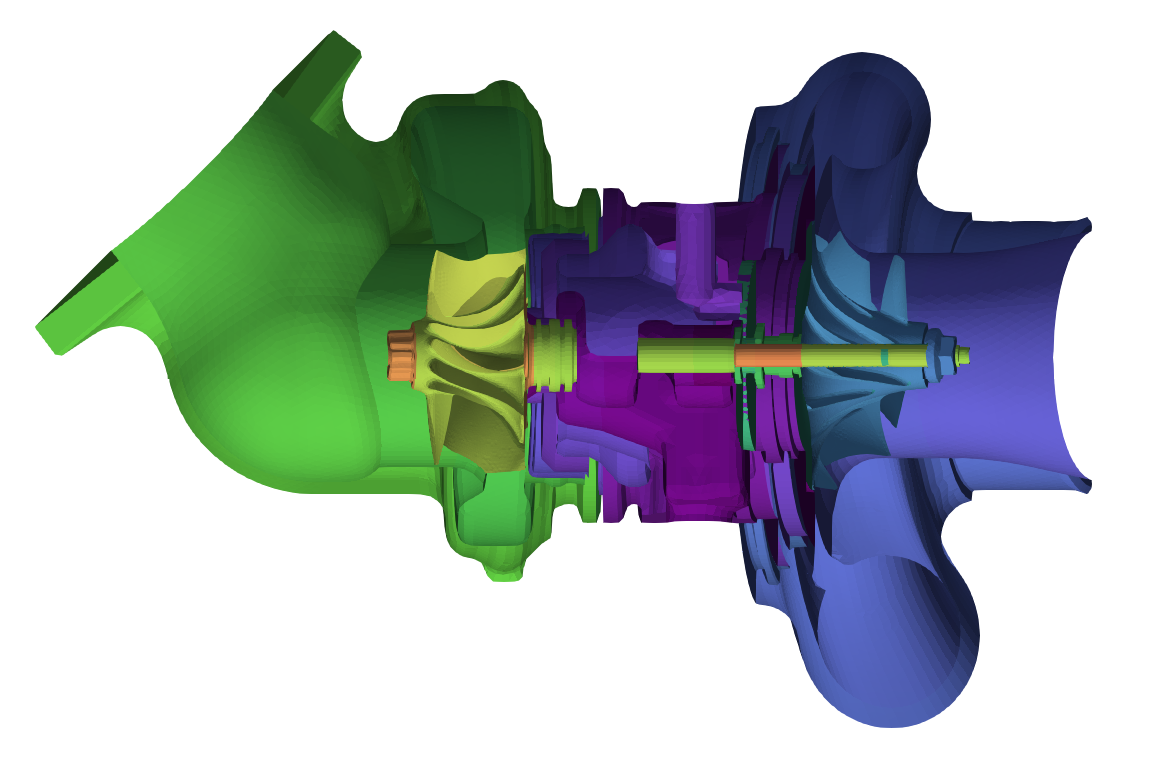}
  \caption{Illustration of the boundary regions on which different
    boundary conditions are applied. This problem has 125 different
    boundary regions.}
  \label{fig:tc_subfacets}
\end{figure}

\begin{figure}
  \centering
  \begin{subfigure}{0.45\textwidth}
    \center\includegraphics[width=0.9\linewidth]{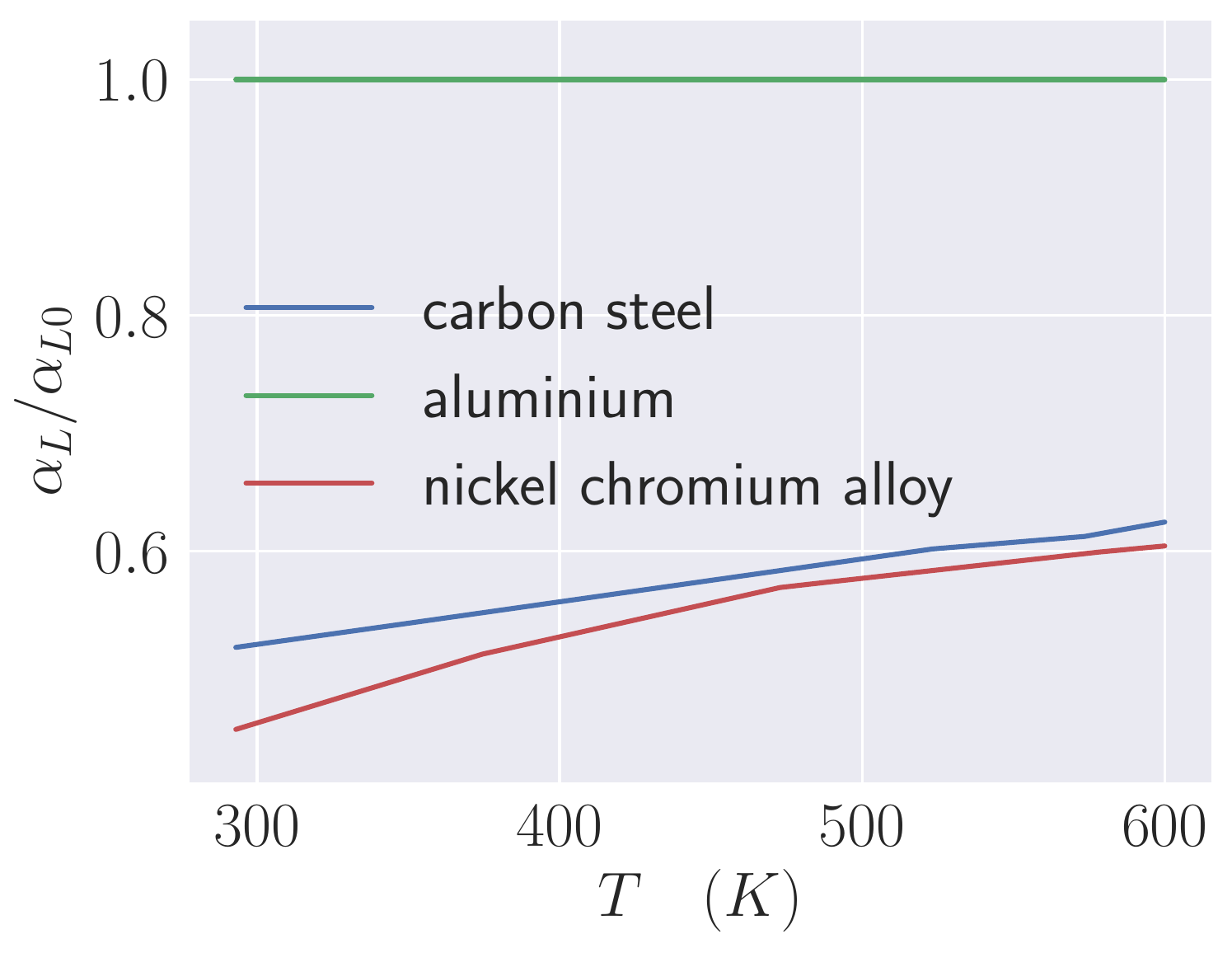}
    \caption{Linear expansivity coefficient.}
  \end{subfigure}
  \begin{subfigure}{0.45\textwidth}
    \center\includegraphics[width=0.9\linewidth]{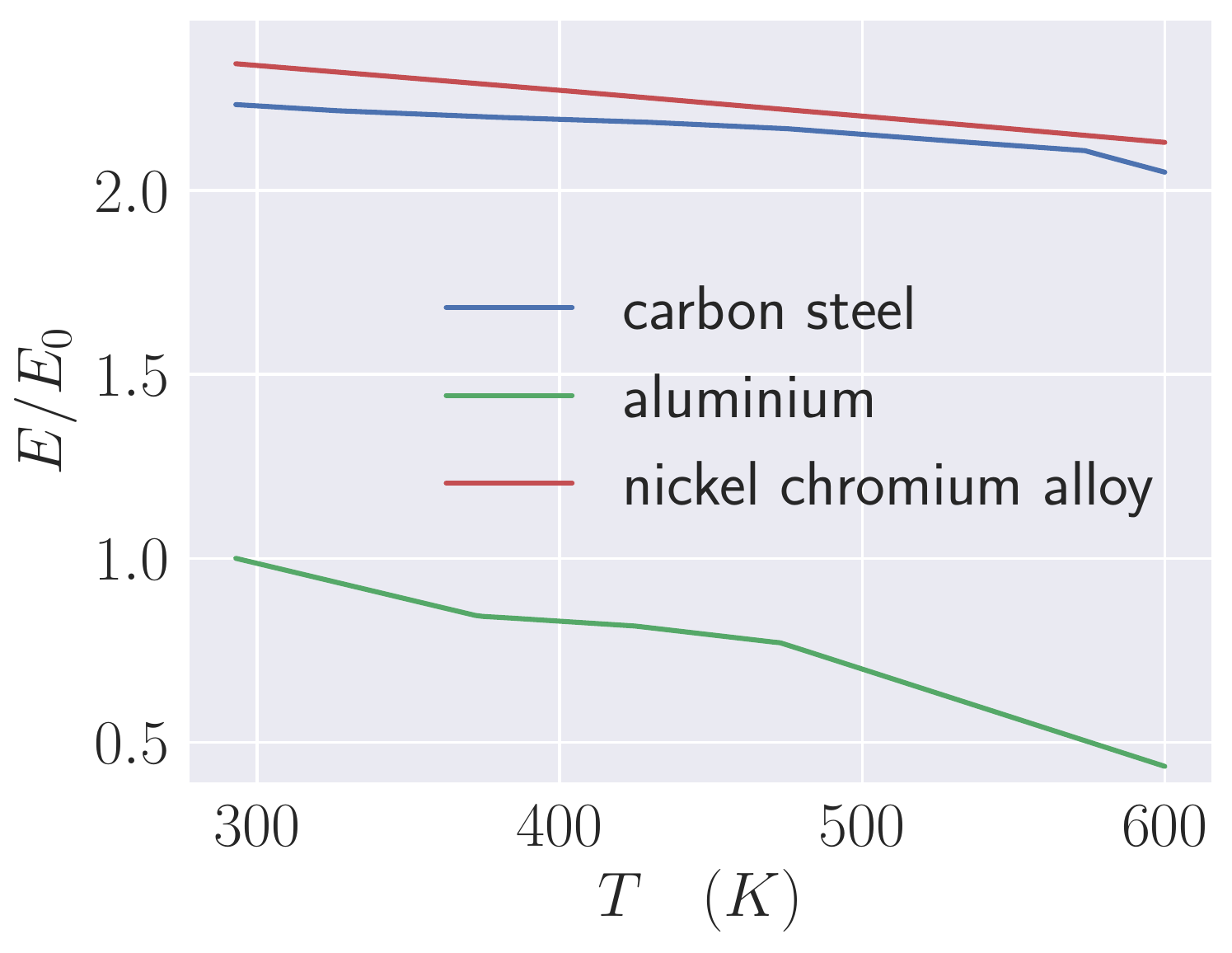}
    \caption{Young's modulus.}
  \end{subfigure}
  \begin{subfigure}{0.45\textwidth}
    \center\includegraphics[width=0.9\linewidth]{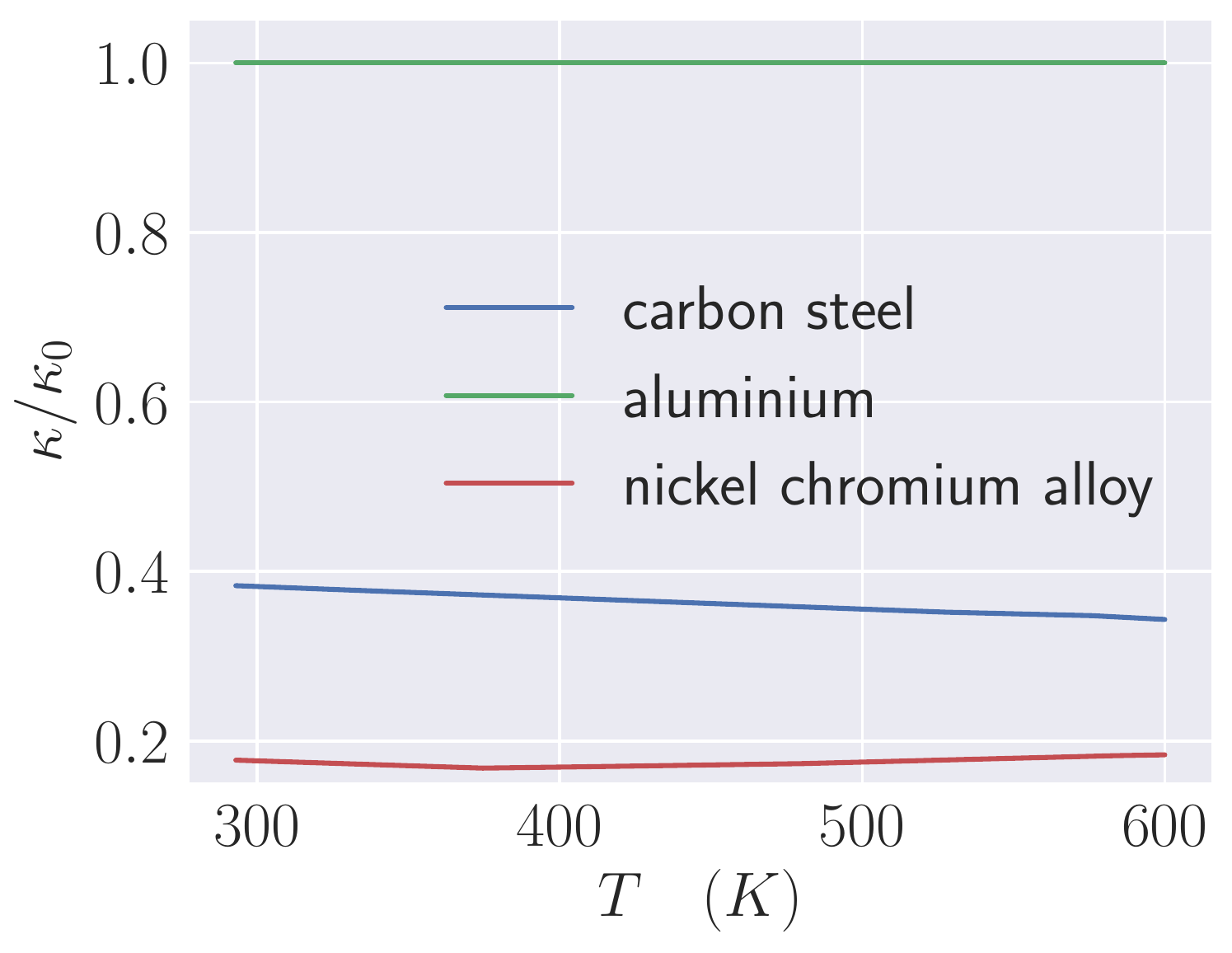}
    \caption{Thermal conductivity.}
  \end{subfigure}
  \begin{subfigure}{0.45\textwidth}
    \center\includegraphics[width=0.9\linewidth]{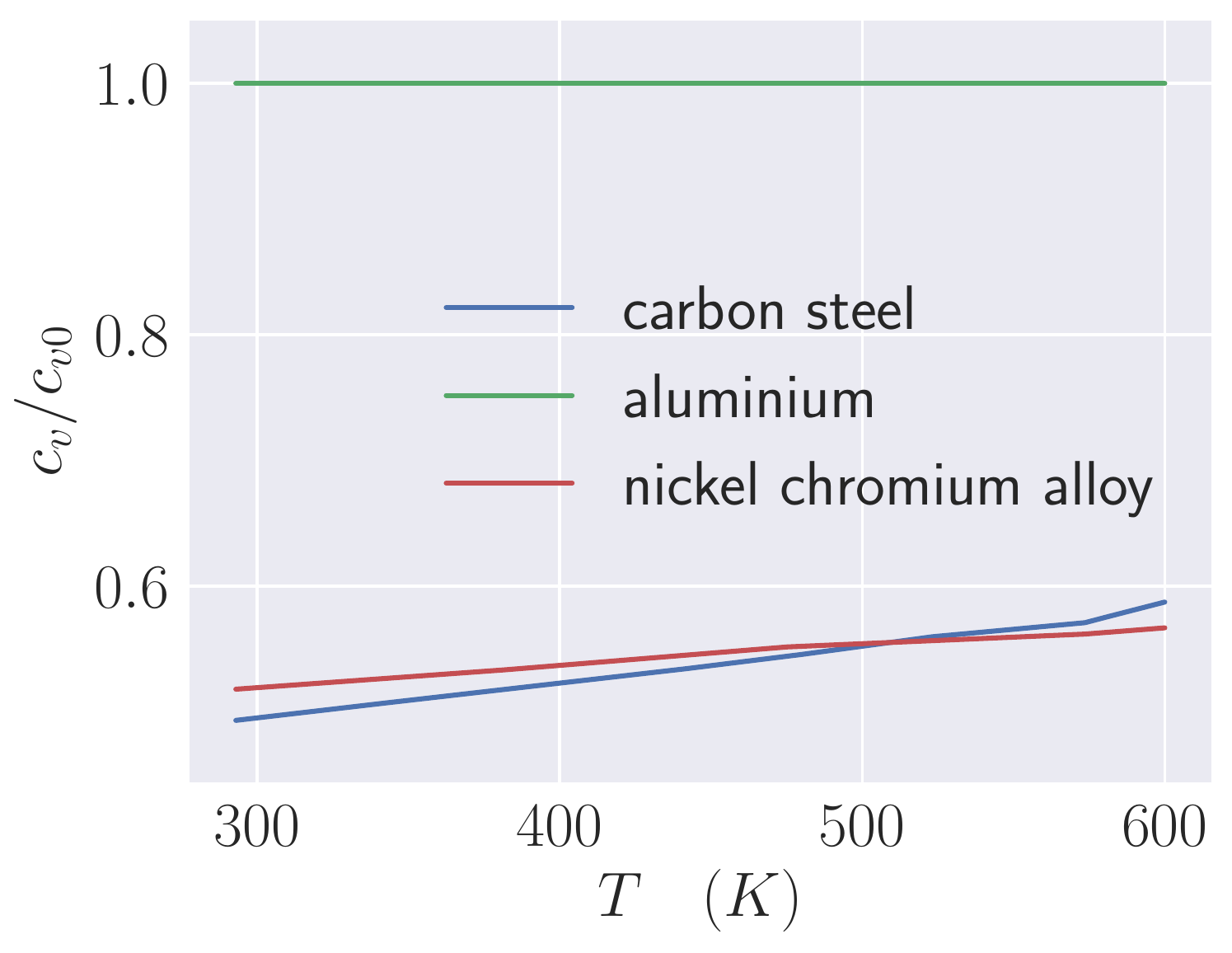}
    \caption{Specific heat.}
  \end{subfigure}
  \caption{Temperature-dependent material data for three of the 12
    materials used in thermomechanical analysis of the turbocharger. The
    data is normalised, where $\alpha_{L0}, E_0, \kappa_0$ and $c_{v0}$
    are the values of $\alpha_{L}, E, \kappa$ and $c_{v}$ at room
    temperature in aluminium.}
  \label{fig:mesh_materials}
\end{figure}

\begin{figure}
  \centering
  \begin{subfigure}{0.45\textwidth}
    \includegraphics[width=0.9\linewidth]{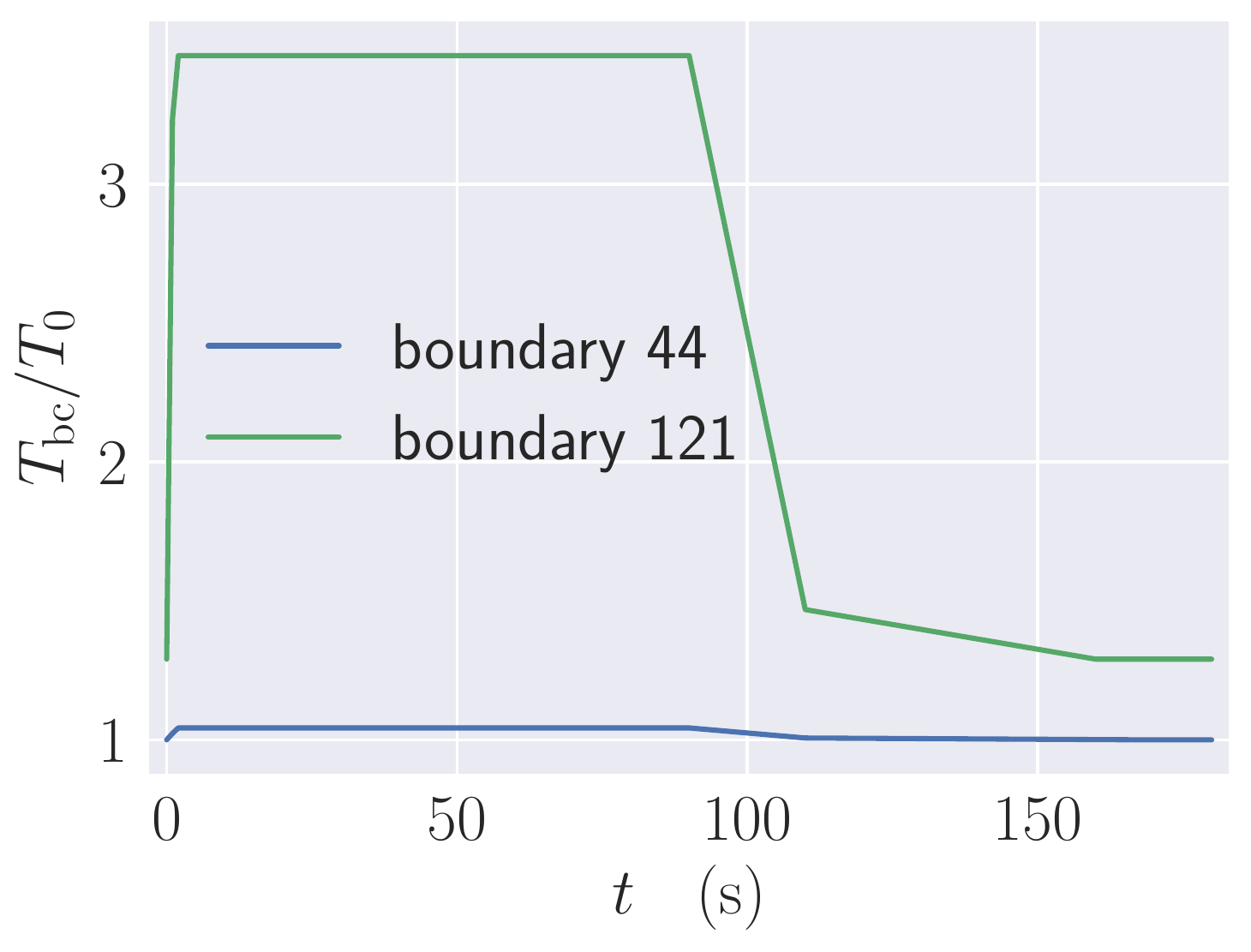}
    \caption{Far field temperature.}
  \end{subfigure}
  \begin{subfigure}{0.45\textwidth}
    \includegraphics[width=0.9\linewidth]{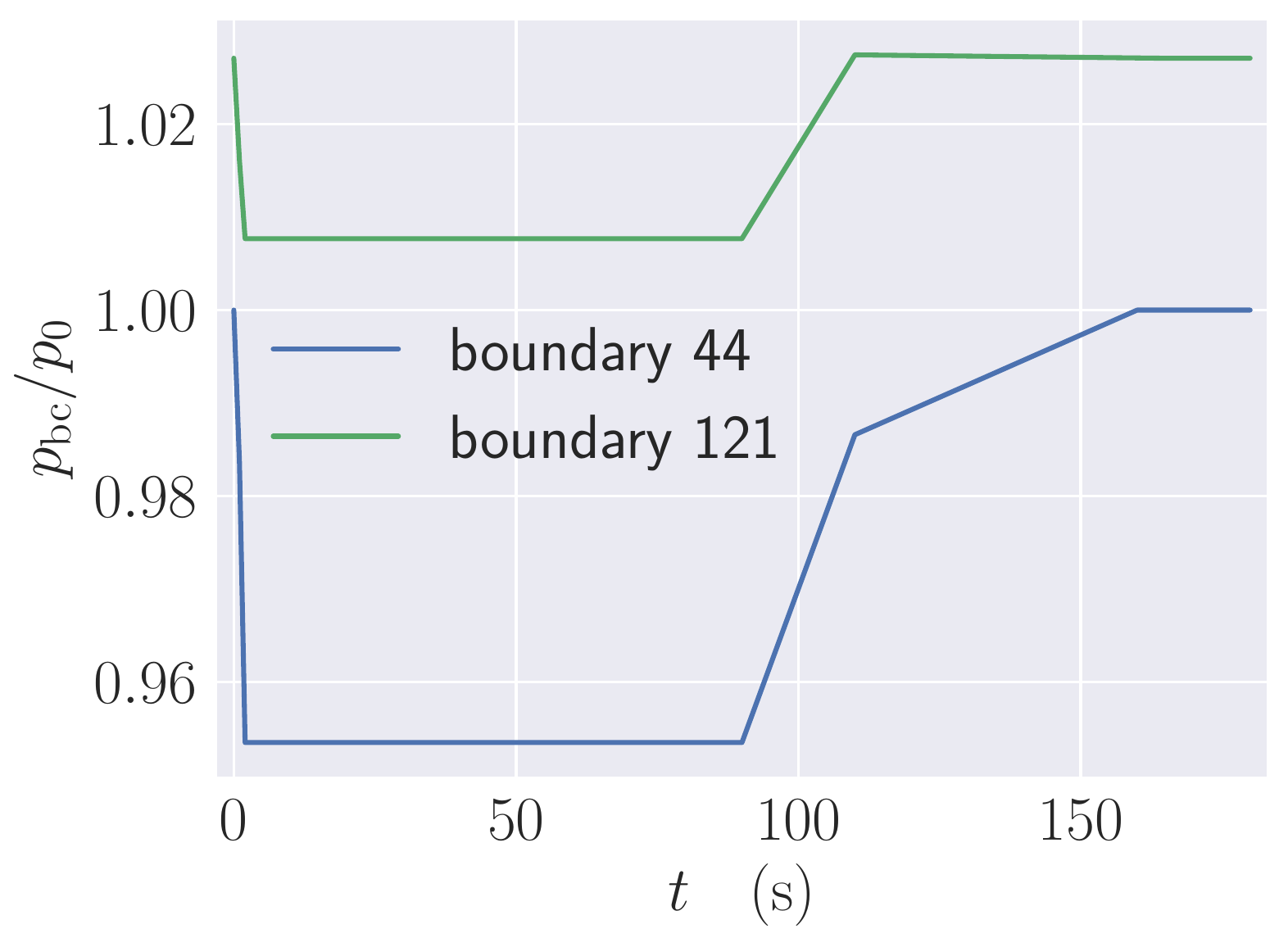}
    \caption{Pressure.}
  \end{subfigure}
  \begin{subfigure}{0.45\textwidth}
    \includegraphics[width=0.9\linewidth]{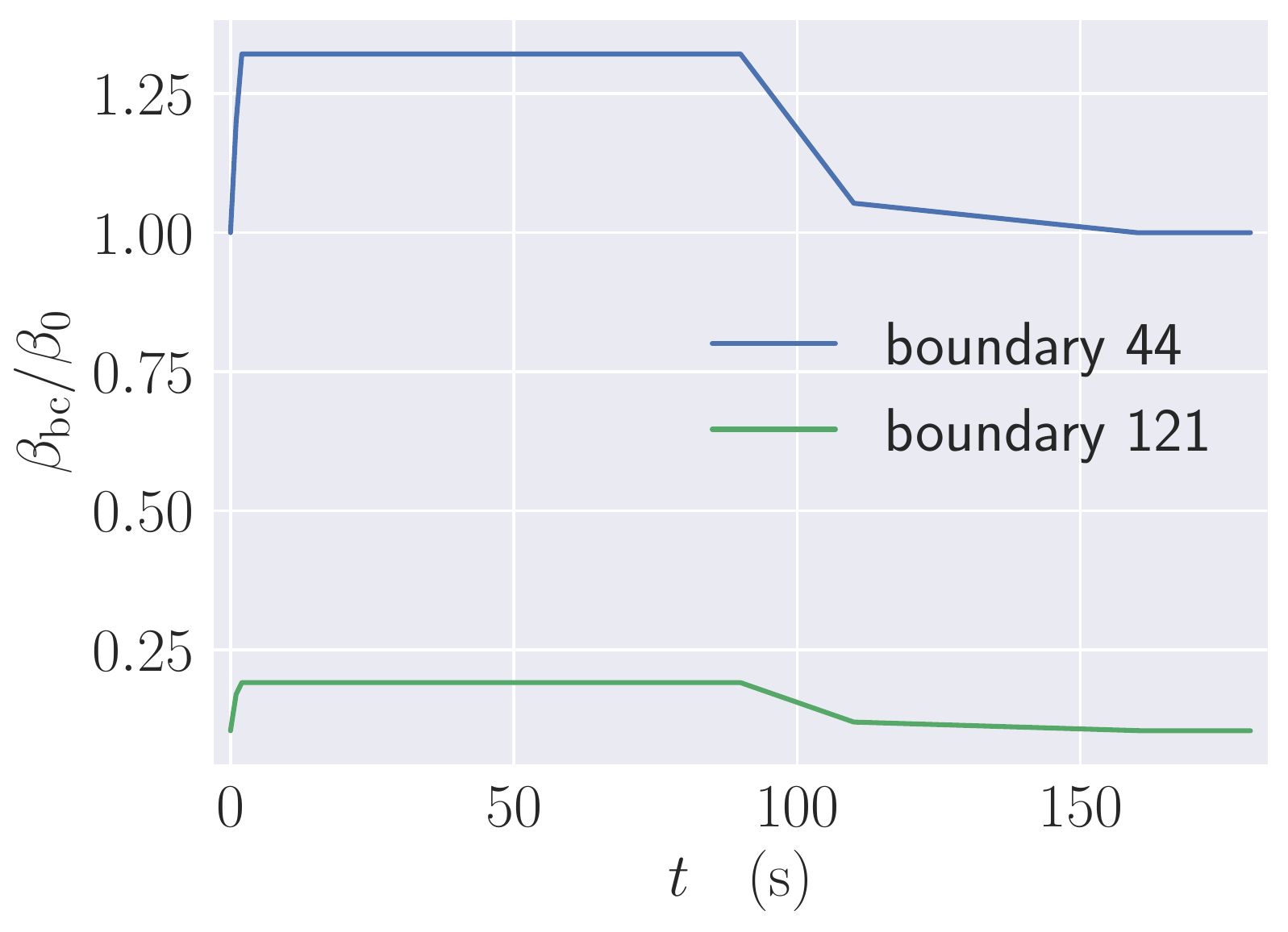}
    \caption{Heat transfer coefficient.}
  \end{subfigure}
  \caption{Time-dependent boundary data for two of the 125 boundary
    conditions used in analysis of the turbocharger. The data is
    normalised, where $T_{0}, p_{0}$ and $\beta_{0}$ are the values of
    $T_{\mathrm{bc}}, p_{\mathrm{bc}}$ and $\beta_{\mathrm{bc}}$ at time
    $t=0$ on boundary~\num{44}.}
  \label{fig:mesh_bcs}
\end{figure}

The test cases are representative of realistic problems in terms of the
number of materials and the number of boundary conditions. The
turbocharger is composed of \num{12} different materials (as illustrated
in \cref{fig:tc_subdomains}) and has 125 boundary regions (as
illustrated in \cref{fig:tc_subfacets}). The steam turbine problem has
\num{8} different materials and \num{135} boundary regions (not shown).
On each boundary region, different time-dependent boundary conditions
are prescribed. Given the complexity of the problems, it is not possible
to fully describe all material properties and boundary conditions. We
therefore provide illustrative examples of typical normalised material
data and boundary conditions that are used. Thermal and elastic
parameters for the different materials are temperature dependent, and
normalised sample material data is shown in \cref{fig:mesh_materials}
for three different materials. \Cref{fig:mesh_bcs} shows time-dependent
boundary condition data for two boundary regions.


%------------------------------------------------------------------------------
\section{Performance studies}
\label{sec:performance_examples}

We present performance data for steady and unsteady cases. We are
primarily interested in total runtime, but also present timing
breakdowns for performance-critical operations, namely: data I/O, matrix
assembly and Newton/linear solver. We focus mainly on steady problems as
the time cost for a steady solution provides an upper bound on the per
time step cost of an unsteady implicit simulation. Simulations were
performed on the UK national supercomputer, ARCHER. ARCHER is a Cray
XC30 system and its key specifications are summarised in
\cref{tab:computer_specs}.

\begin{table}
  \centering
  \begin{tabular}{l|l}
    Processors (per node) & 2 $\times$ Intel E5-2697 v2 \\
    Cores per node        & \num{24}                    \\
    Clock speed           & \SI{2.70}{\giga\hertz}      \\
    Memory per node       & \SI{64}{\giga\byte}         \\
    Interconnect          & Cray Aries                  \\
    Filesystem            & Lustre
  \end{tabular}
  \caption{Overview of the ARCHER Cray XC30 system.}
  \label{tab:computer_specs}
\end{table}

%------------------------------------------------------------------------------
\subsection{Solver configuration}
\label{sec:solver_configuration}

All tests use the conjugate gradient method preconditioned with
algebraic multigrid (AMG). We stress that AMG is not a black-box method;
our experience is that simulation time is often poor and can fail with
default settings, especially for the mechanical problem. For the thermal
solve, we precondition using BoomerAMG~\citep{henson:2002} from the
HYPRE library, which is a classical AMG implementation. For the elastic
solve, we precondition using GAMG~\citep{adams:gamg}, the native PETSc
smoothed aggregation AMG implementation. Classical AMG tends to be best
suited to scalar-valued equations, and smoothed aggregation is suited to
vector-valued equations.

With BoomerAMG, we use a hybrid Gauss--Seidel smoother from the HYPRE
library~\cite{Baker2011}. With GAMG, we use a Chebyshev smoother. It is
important for Chebyshev smoothing that the maximum eigenvalues are
adequately approximated. If they are not, the preconditioned system may
lose positive-definiteness and the CG methods will therefore fail. We
have observed for complicated geometries and for higher-order elements
that more Krylov iterations are typically required (more than the GAMG
default) to adequately estimate the highest eigenvalues compared to
simple geometries. It can also be important to control the rate of
coarsening, particularly for linear tetrahedral elements, when using
smoothed aggregation. For good performance, it is recommended to ensure
that the multigrid preconditioner coarsens sufficiently quickly. For
complicated geometries where robustness is an issue (typically on lower
quality meshes), we have observed heuristically that increasing the size
of the `coarse grid' used by the multigrid preconditioner (the level at
which the preconditioner ceases to further coarsen the algebraic
problem) for the mechanical problem can dramatically enhance robustness,
especially on lower quality meshes. The scalar thermal solve is
typically more robust and requires fewer iterations than the
vector-valued mechanical problem.

%------------------------------------------------------------------------------
\subsection{Steady thermomechanical simulations}

For steady simulations, the nonlinear thermal problem is first solved,
followed by the temperature-dependent mechanical problem.  The thermal
model parameters are temperature dependent, so an initial guess for the
temperature field is required. For all examples, the initial guess of
the temperature field is \SI{400}{\kelvin} and the Newton solver is
terminated once a relative residual of \num{e-9} is reached. The
iterative solver for the mechanical problem is terminated once a
preconditioned relative residual norm of~\num{e-6} is reached.

%------------------------------------------------------------------------------
\subsubsection{Turbocharger}

\begin{figure}
  \centering
  \begin{subfigure}{0.49\textwidth}
    \centering
    \includegraphics[width=\textwidth]{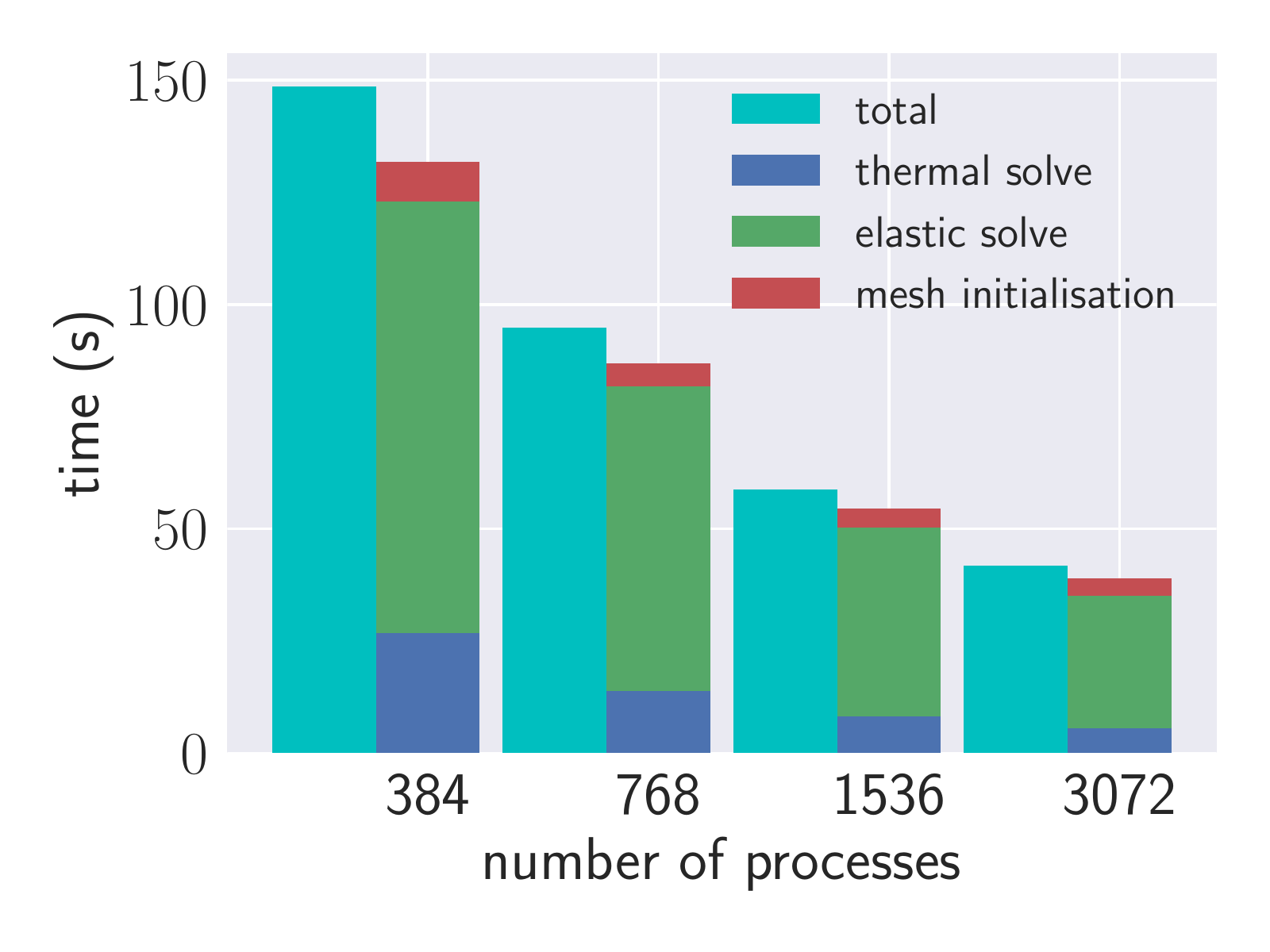}
    \caption{Polynomial degree $p = 1$. }
    \label{fig:tc_steady_te_strong_scl_archer_p1}
  \end{subfigure}
  \begin{subfigure}{0.49\textwidth}
    \centering
    \includegraphics[width=\textwidth]{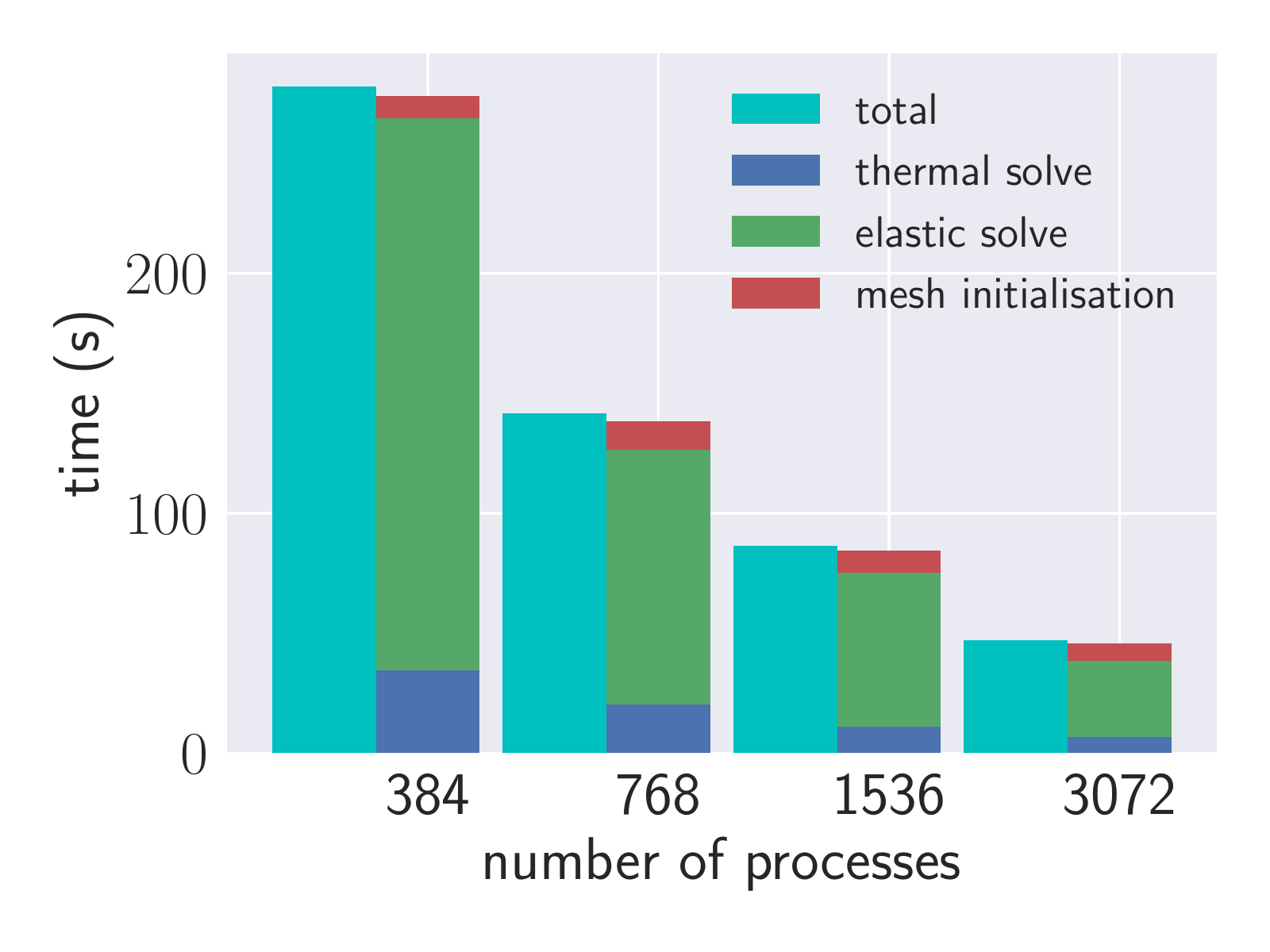}
    \caption{Polynomial degree $p = 2$.}
    \label{fig:tc_steady_te_strong_scl_archer_p2}
  \end{subfigure}
  \caption{Strong scaling results for the steady turbocharger problem
  using mesh~`B'. For both cases the thermal and elastic problems have
  \num{67373812} and \num{202121436} degrees-of-freedom, respectively.
  The mesh used for the $p=1$ case has been refined uniformly once.}
  \label{fig:tc_steady_te_strong_scl}
\end{figure}

\Cref{fig:tc_steady_te_strong_scl} presents strong scaling results for
the turbocharger using linear ($p = 1$) and quadratic elements ($p =
2$). Both cases have the same number of degrees-of-freedom -- over 67~M
for the temperature field and over 202~M for the displacement field.
Also shown are breakdowns of the time cost for keys steps: reading and
partitioning the mesh, and solving the thermal and elastic problems. We
see that the scaling trend is good, the wall-clock time is very low in
view of the problem sizes, and that the elastic solve is the dominant
cost. We present the timings using a linear wall-clock time scale to
make clear the potential impact on real design processes. Interpreting
the timing between two process counts should be done with some caution
because with unstructured grids the distribution of the mesh and the
aggregation created by the multigrid implementations will differ when
changing the number of processes.

\begin{figure}
  \centering
  \begin{subfigure}{0.5\textwidth}
    \centering
    \includegraphics[height=0.3\textheight]{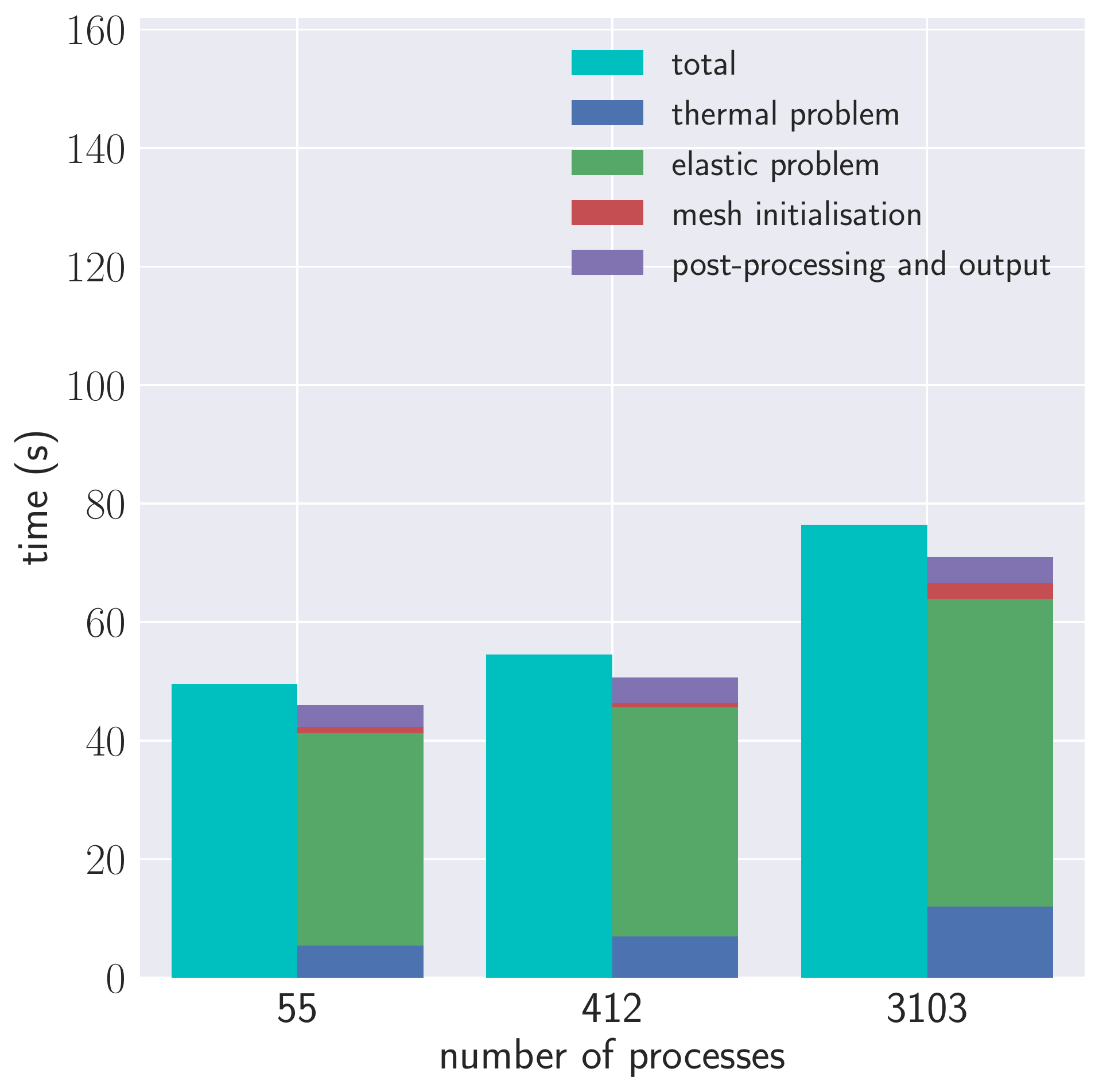}
    \caption{Turbocharger mesh `A'.}
  \end{subfigure}%
  \begin{subfigure}{0.37\textwidth}
    \centering
    \includegraphics[height=0.3\textheight]{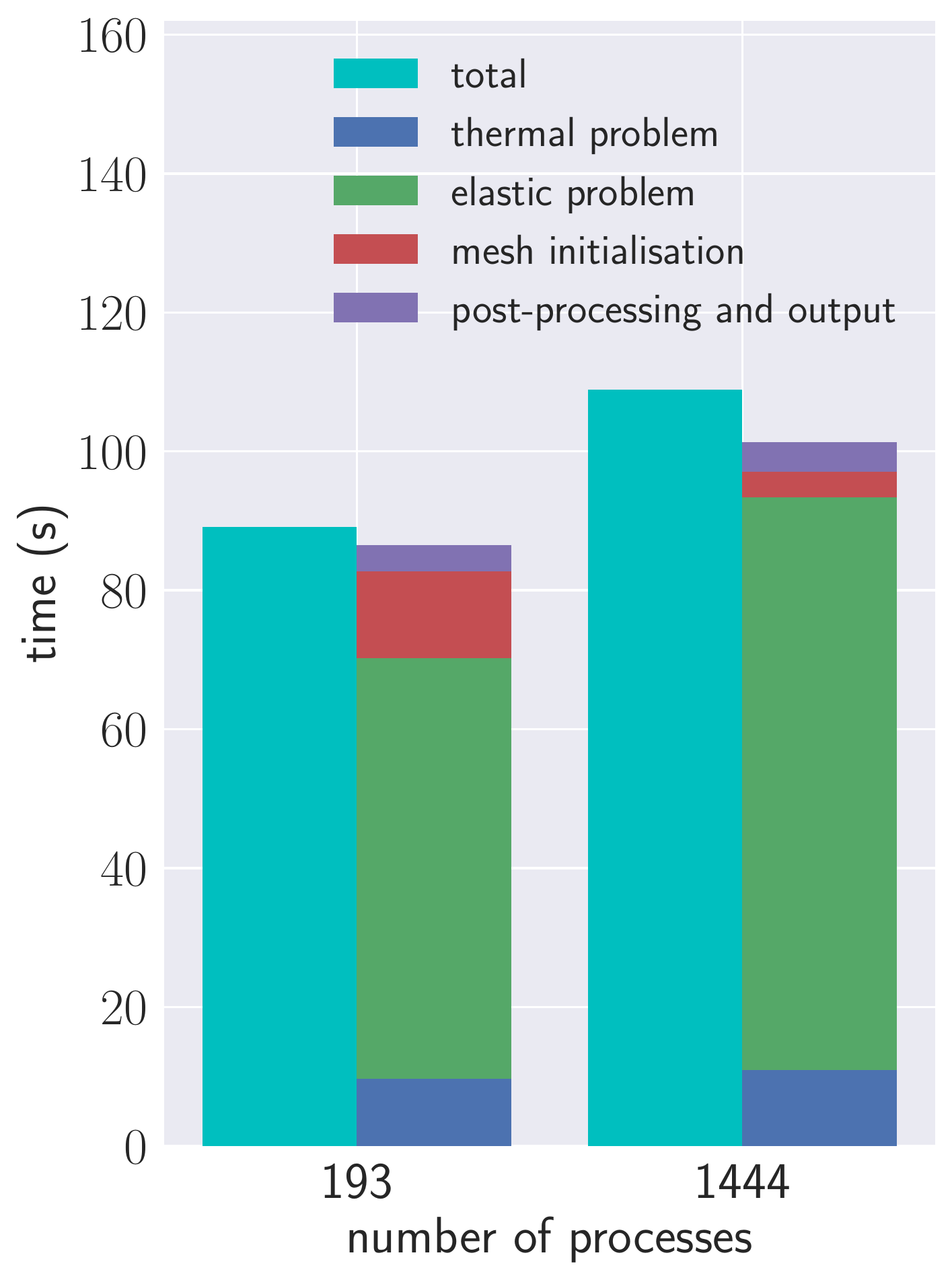}
    \caption{Turbocharger mesh `B'.}
  \end{subfigure}
  \caption{Weak scaling results for the steady turbocharger problem
  using $p = 1$ elements. The number of displacement degrees-of-freedom
  is kept close to
	\num{1.4e5}~per process.}
  \label{fig:tc_weak_scaling}
\end{figure}

Weak scaling results are presented in \cref{fig:tc_weak_scaling} for
linear elements, with approximately \num{1.4e5} displacement
degrees-of-freedom per process. The thermal problem size ranges from
\num{2566244} to \num{143630023} degrees-of-freedom, and the elastic
problem from \num{7698732} to \num{430890069} degrees-of-freedom. We
observe satisfactory weak scaling.

\begin{figure}
  \centering
  \includegraphics[width=0.95\textwidth]{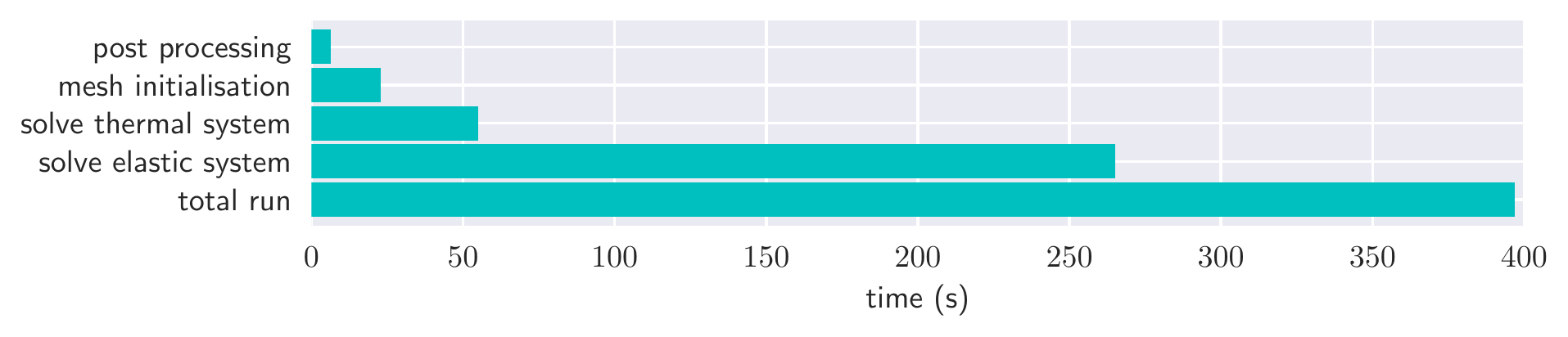}
  \caption{Runtime breakdown for the steady-state turbocharger problem
    with \num{1121793507} thermal and \num{3365380521} elastic
    degrees-of-freedom  using \num{24576} MPI processes.}
  \label{fig:three_billion_dof_tc}
\end{figure}

To demonstrate the potential for solving extreme scale problems, the
turbocharger problem has been solved with over \SI{3.3e9} displacement
degrees-of-freedom using quadratic ($p = 2$) elements. The mesh was
generated by refining turbocharger mesh~`A' recursively three times. The
resulting mesh has \num{820960256} cells and \num{163283303} vertices.
The simulation was run using \num{24576}~MPI processes, and the
time-to-solution was under \SI{400}{\second}. A breakdown of the timings
is presented in \cref{fig:three_billion_dof_tc}. The computation time is
dominated by the elastic solve, taking \SI{67}{\percent} of the total
runtime, with \SI{14}{\percent} of the time spent on the thermal solve.
The example shows that there is potential to move well beyond current
limits on problem size for thermomechanical simulation.

%------------------------------------------------------------------------------
\subsubsection{Steam turbine}

\begin{figure}{}
  \centering
  \begin{subfigure}[t]{0.49\textwidth}
    \centering
    \includegraphics[height=0.225\textheight]{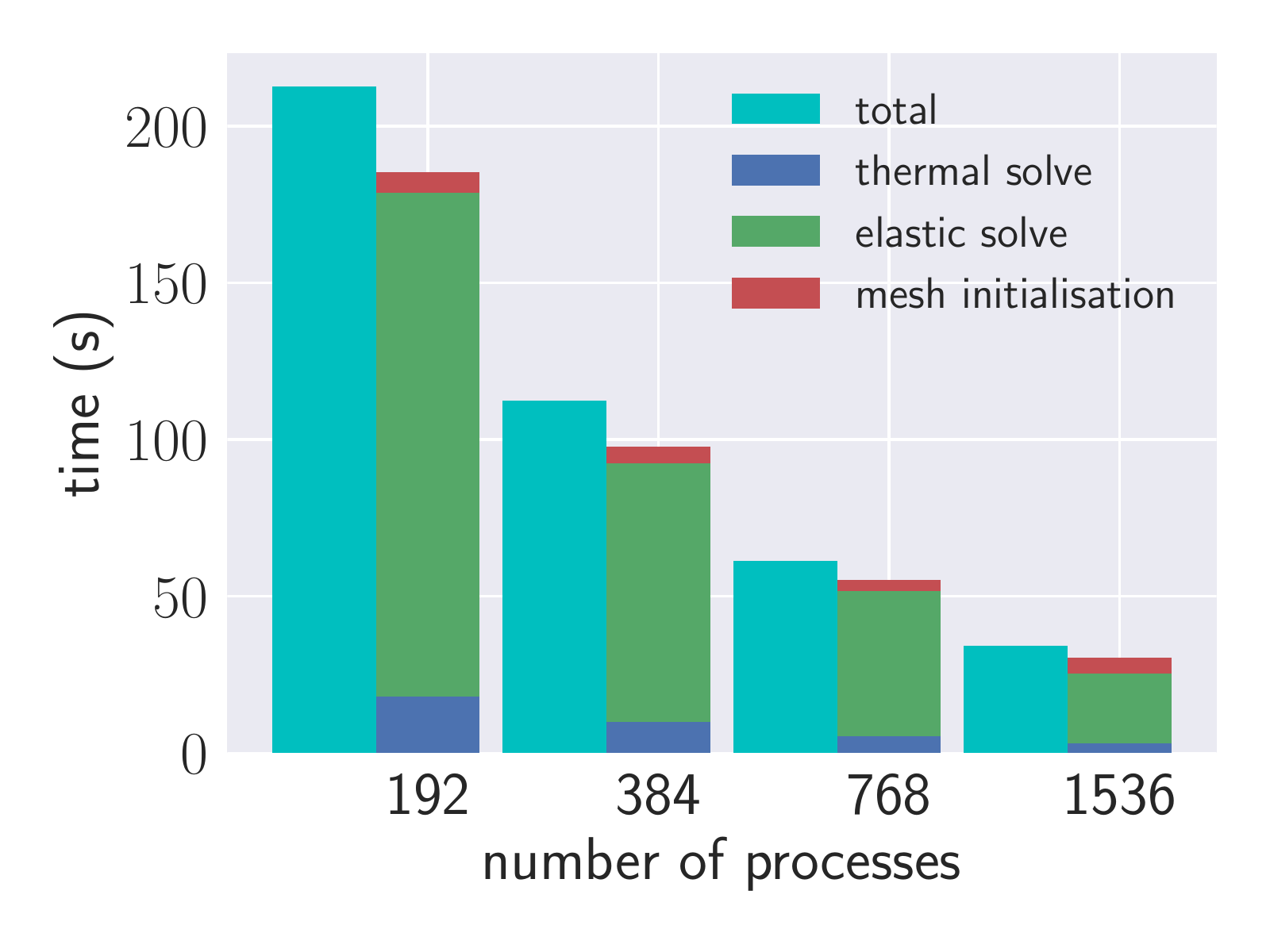}
    \caption{Polynomial degree $p = 1$ with one level of refinement.}
    \label{fig:st_steady_te_strong_scl_archer}
  \end{subfigure}
  \begin{subfigure}[t]{0.49\textwidth}
    \centering
    \includegraphics[height=0.225\textheight]{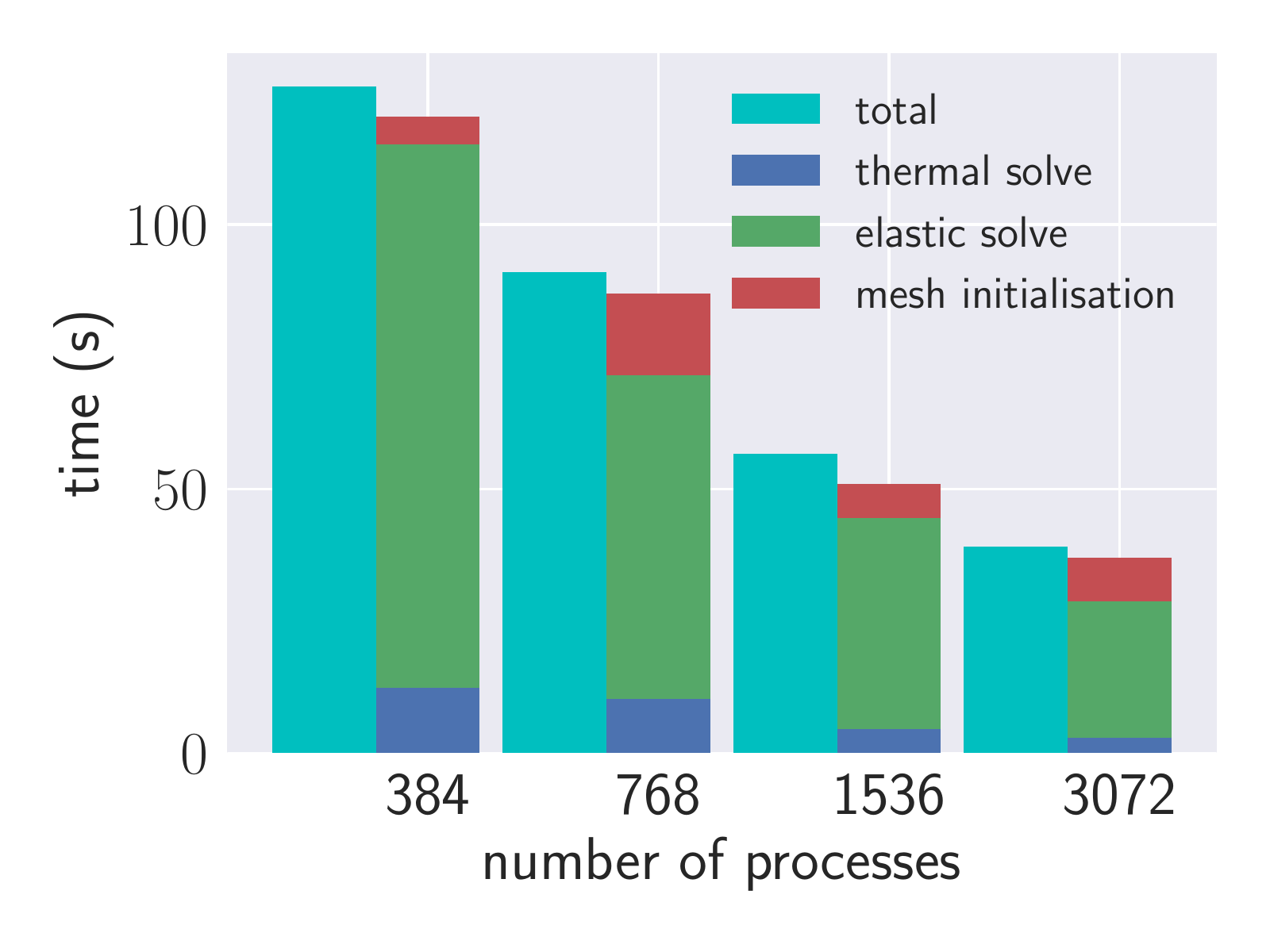}
    \caption{Polynomial degree $p = 2$.}
    \label{fig:st_steady_te_p2_strong_scl_archer}
  \end{subfigure}
  \caption{Timings for the steady-state steam turbine problem. Both
  cases have \num{36325419} thermal  degrees-of-freedom and
  \num{108976257} elastic degrees-of-freedom.}
  \label{fig:st_steady_te_strong_scl}
\end{figure}

A distinguishing feature of the steam turbine problem compared to the
turbocharger is the presence of more fine (slender) geometric features
and lower mesh quality (see \cref{fig:mesh_quality}).
\Cref{fig:st_steady_te_strong_scl} presents strong scaling results for
the steam turbine with over 36~M thermal and over 108~M displacement
degrees-of-freedom (linear and quadratic elements). The runtimes are
good, and the scaling satisfactory, as observed for the turbocharger.
With \num{3072} MPI processes this gives an average of \num{35474}
degrees-of-freedom per process. We observe that as the degree-of-freedom
per process count reduces below \num{30000}, the solution time suffers
and does not scale well, which is likely due to the dominance of
inter-process communication.

\begin{figure}
  \centering
  \includegraphics[height=0.25\textheight]{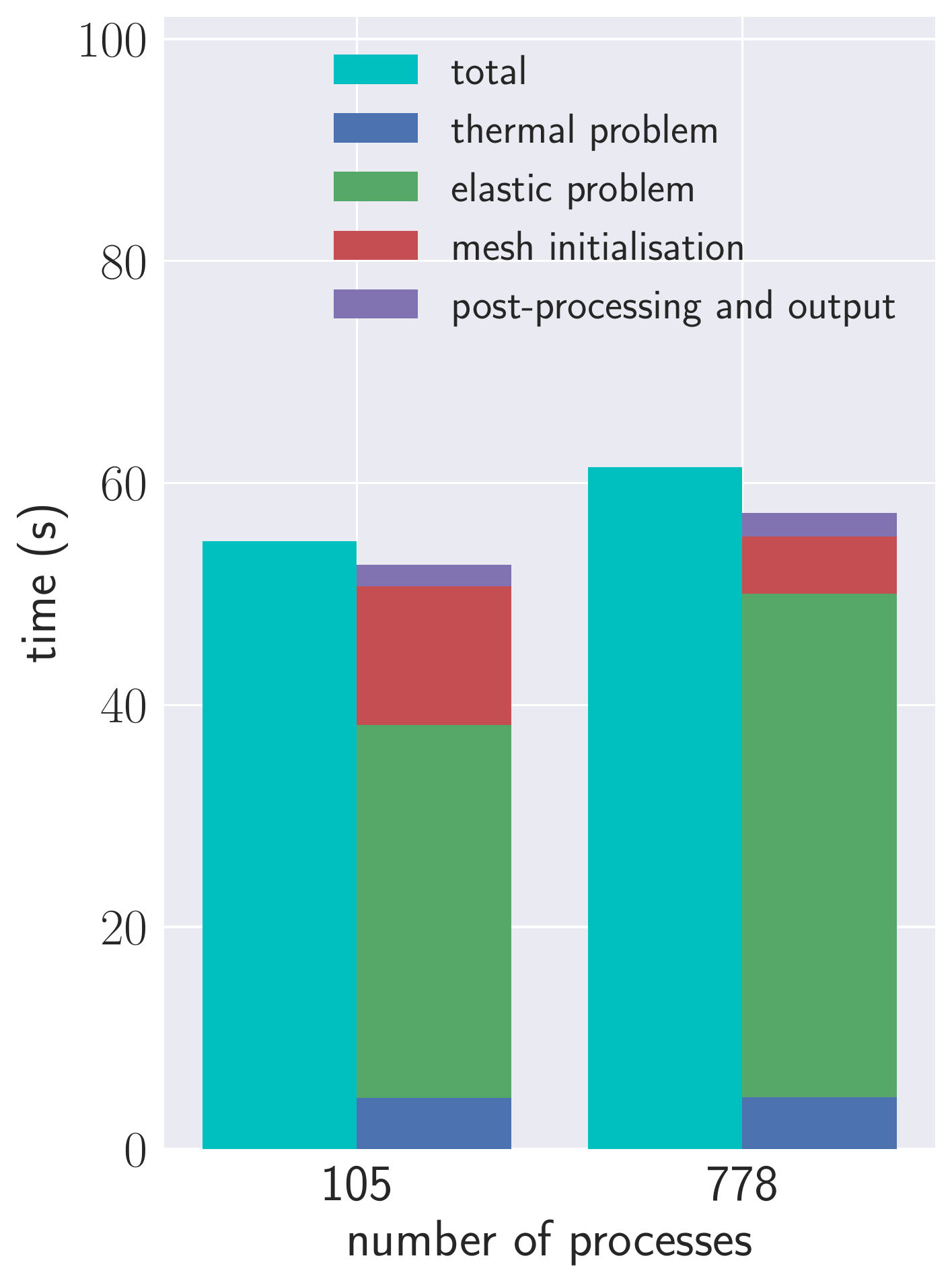}
  \caption{Weak scaling of the thermomechanical analysis of the steam
    turbine with~$p=1$. The number of displacement degrees-of-freedom
    per process is kept close to~\num{1.4e5}.}
  \label{fig:st_weak_scaling}
\end{figure}

Two data points for weak scaling are shown in
\cref{fig:st_weak_scaling}, where the number of displacement
degrees-of-freedom per process is kept close to~\num{1.4e5}. The coarse
problem has \num{4918704} thermal and \num{14756112} displacement
degrees-of-freedom, and the fine problem has \num{36325419} thermal and
\num{108976257} displacement degrees-of-freedom. As with the preceding
results, we see that the elastic solver cost is dominant, and the most
challenging to scale.

%------------------------------------------------------------------------------
\subsection{Transient thermomechanical simulations}

We consider an unsteady simulation of the turbocharger. The steady
performance results provide a good upper bound on the per time step cost
of the implicit unsteady simulations. The per time step cost of an
unsteady simulation will generally be lower than for the steady case
because preconditioners can be re-used, and in many cases the solution
from the previous step can be used as an initial guess for the iterative
solver.

The transient examples correspond to a test cycle in the case of the
turbocharger, and a start-up procedure to operating temperature in the
case of the steam turbine casing. For both problems the initial
temperature is set to \SI{293}{\kelvin}. The transient response is then
driven by time-dependent temperature and pressure boundary conditions.
The time step is adjusted adaptively to limit the maximum temperature
change at any point in the domain to \SI{10}{\kelvin} using
\cref{eq:dt_spec}. Transient simulations use the backward Euler method
($\theta = 1$).

\begin{figure}
  \centering
  \begin{subfigure}{0.5\textwidth}
    \centering
    \includegraphics[width=\textwidth]{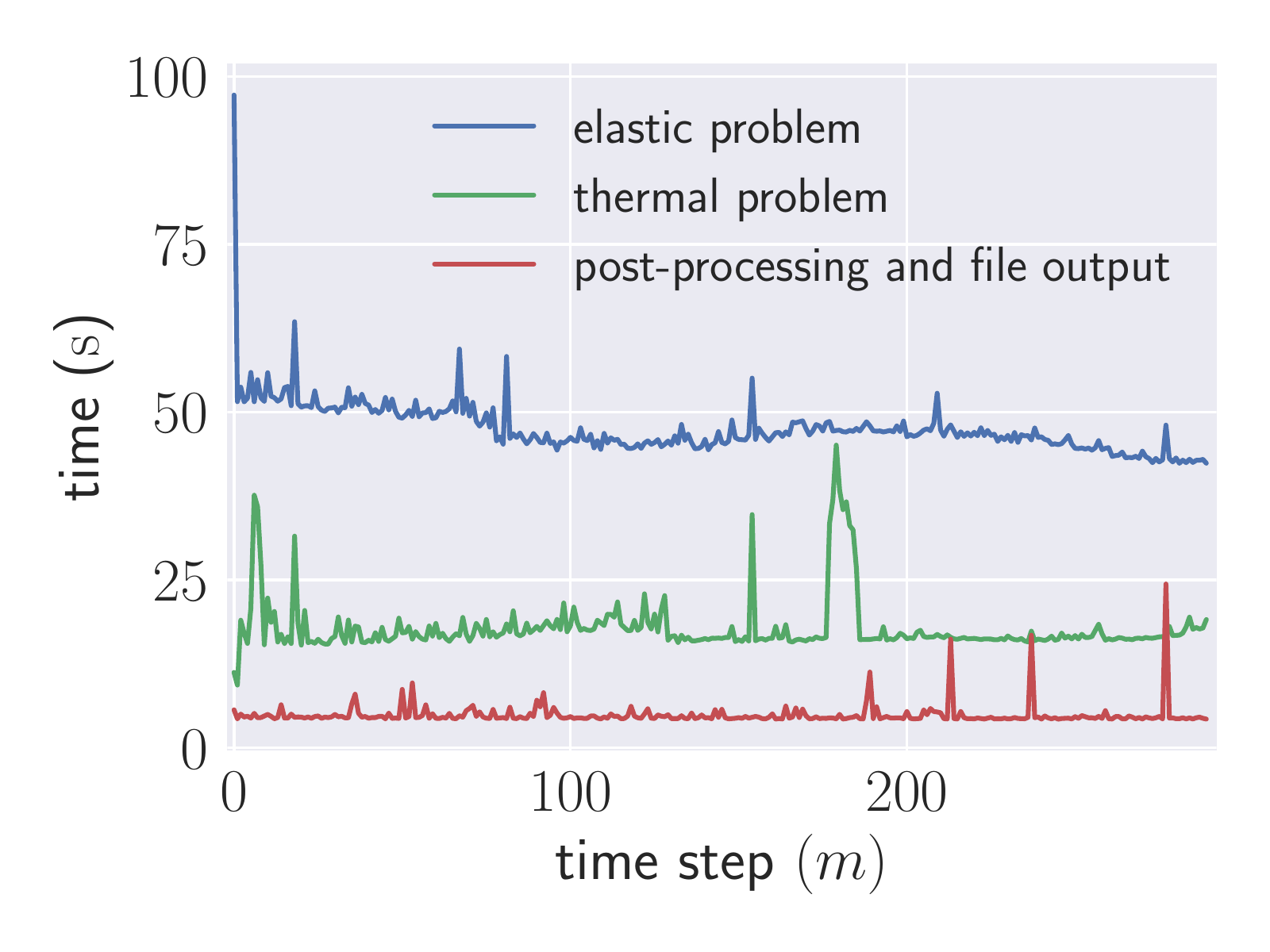}
    \caption{Elastic, thermal and post-processing time.}
  \end{subfigure}

  \begin{subfigure}{0.45\textwidth}
    \centering
    \includegraphics[width=\textwidth]{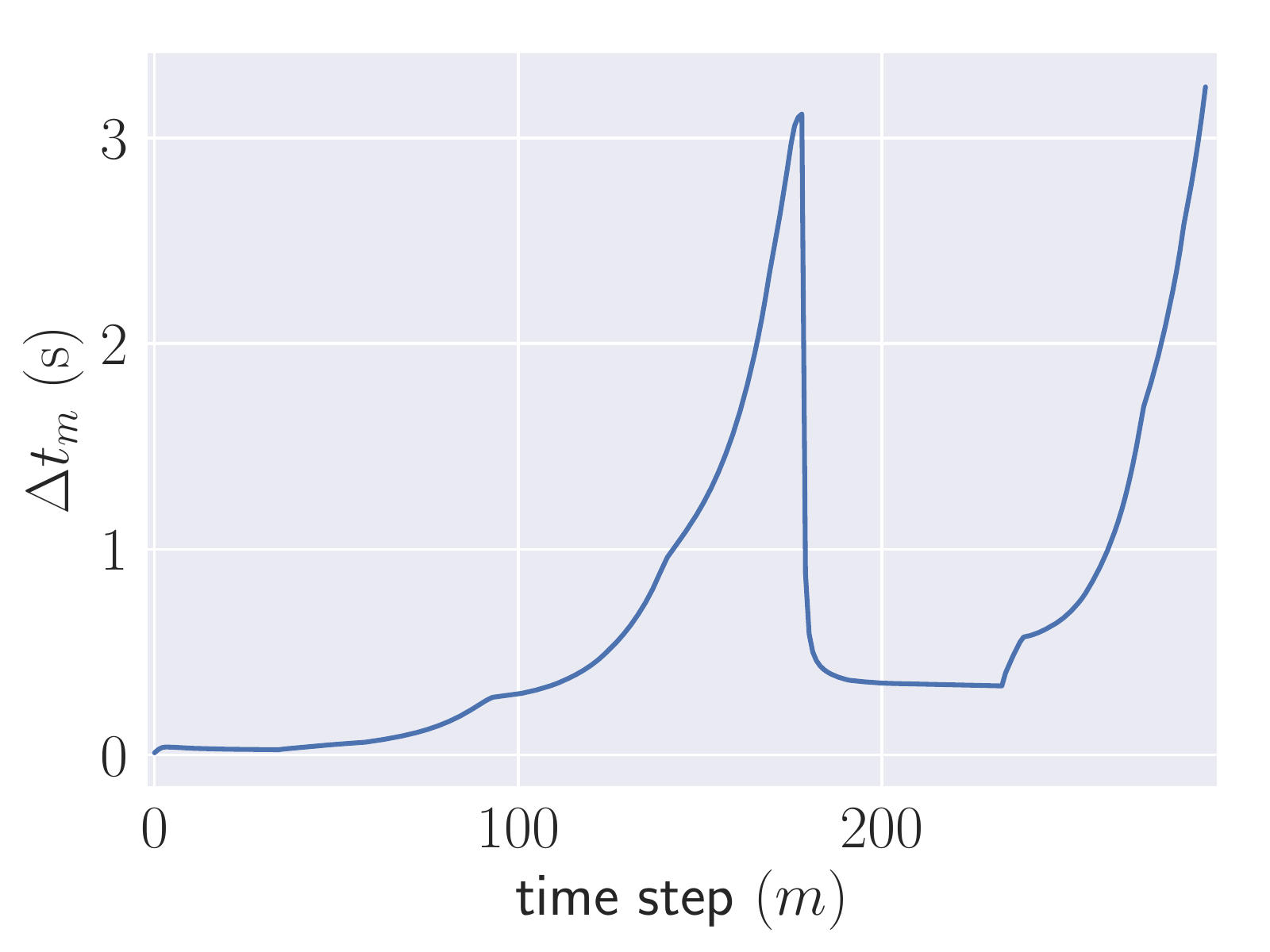}
    \caption{Adaptive time step.} \label{fig:subfig:tc_transient_dt}
  \end{subfigure}
  \begin{subfigure}{0.45\textwidth}
    \centering
    \includegraphics[width=\textwidth]{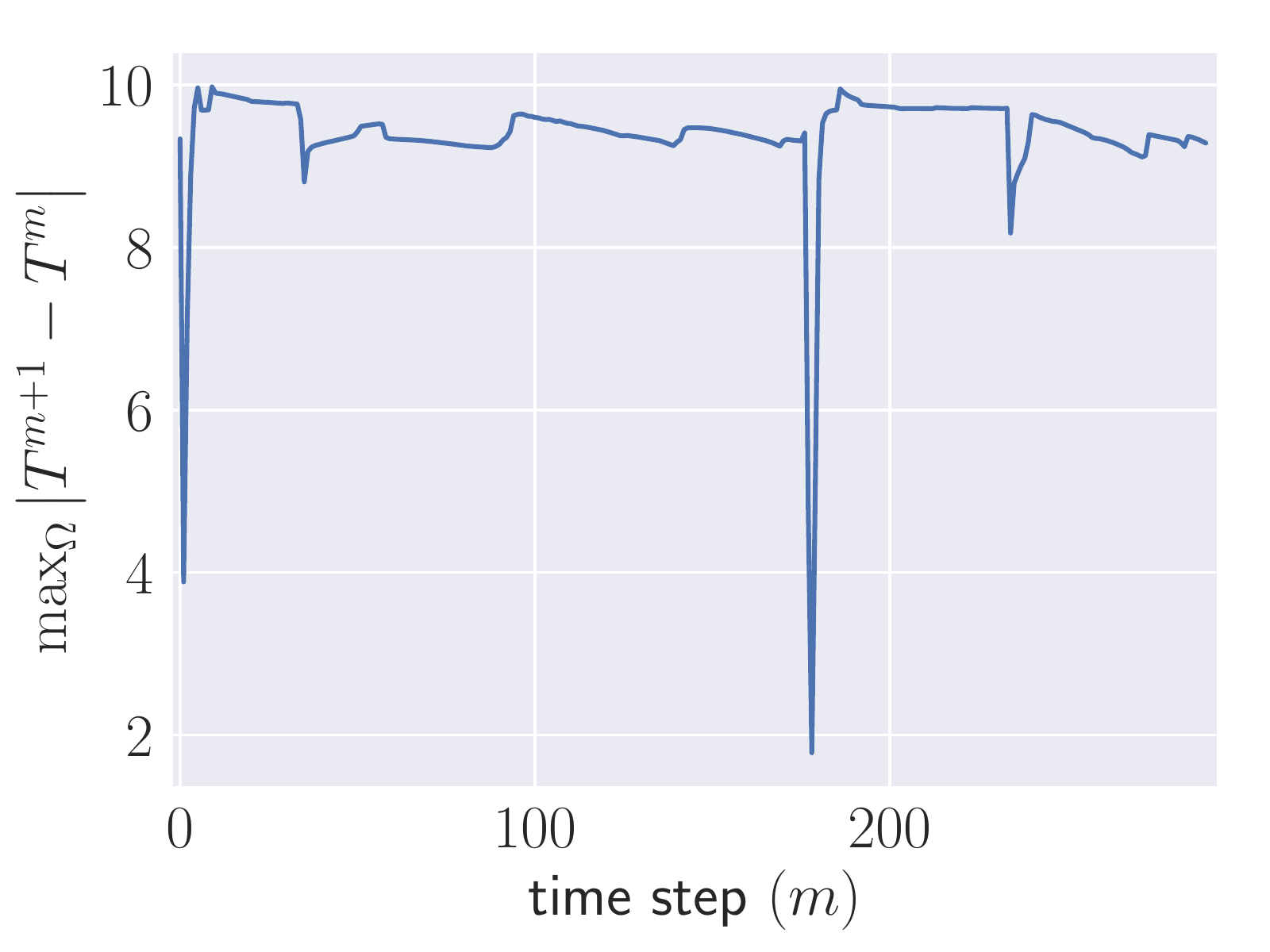}
    \caption{Maximum change in the temperature.}
  \end{subfigure}
  \caption{Simulation data for the transient analysis of the
    turbocharger over one cycle of operation using turbocharger mesh~`B'
    and $p=2$.}
  \label{fig:tc_transient_details}
\end{figure}

A test cycle for the turbocharger problem using mesh `B' with
\num{67373812} thermal and \num{202121436} elastic degrees-of-freedom
($p = 2$ for temperature and displacement fields) was computed. The
simulation required \num{289} time steps to complete a cycle with the
\SI{10}{\kelvin} limit on the maximum temperature change per time step.
The runtime for the simulation was \SI{314}{\minute} using \num{768} MPI
processes. The cost per time step for different components of the
simulation are shown in \cref{fig:tc_transient_details}, as well as the
$\Delta t$ per step and the maximum change in temperature per step. The
cost for the elastic solve is highest for the first step, as the
preconditioner for the elastic part of the problem is re-used for
subsequent time steps. Spikes in the solve time for the thermal problem
correspond to steps at which the thermal preconditioner is rebuilt, and
correspond to times at which rapid temperature changes occur.

\begin{figure}
  \centering
  \begin{subfigure}[t]{0.49\textwidth}
    \centering
    \includegraphics[height=0.25\textheight]{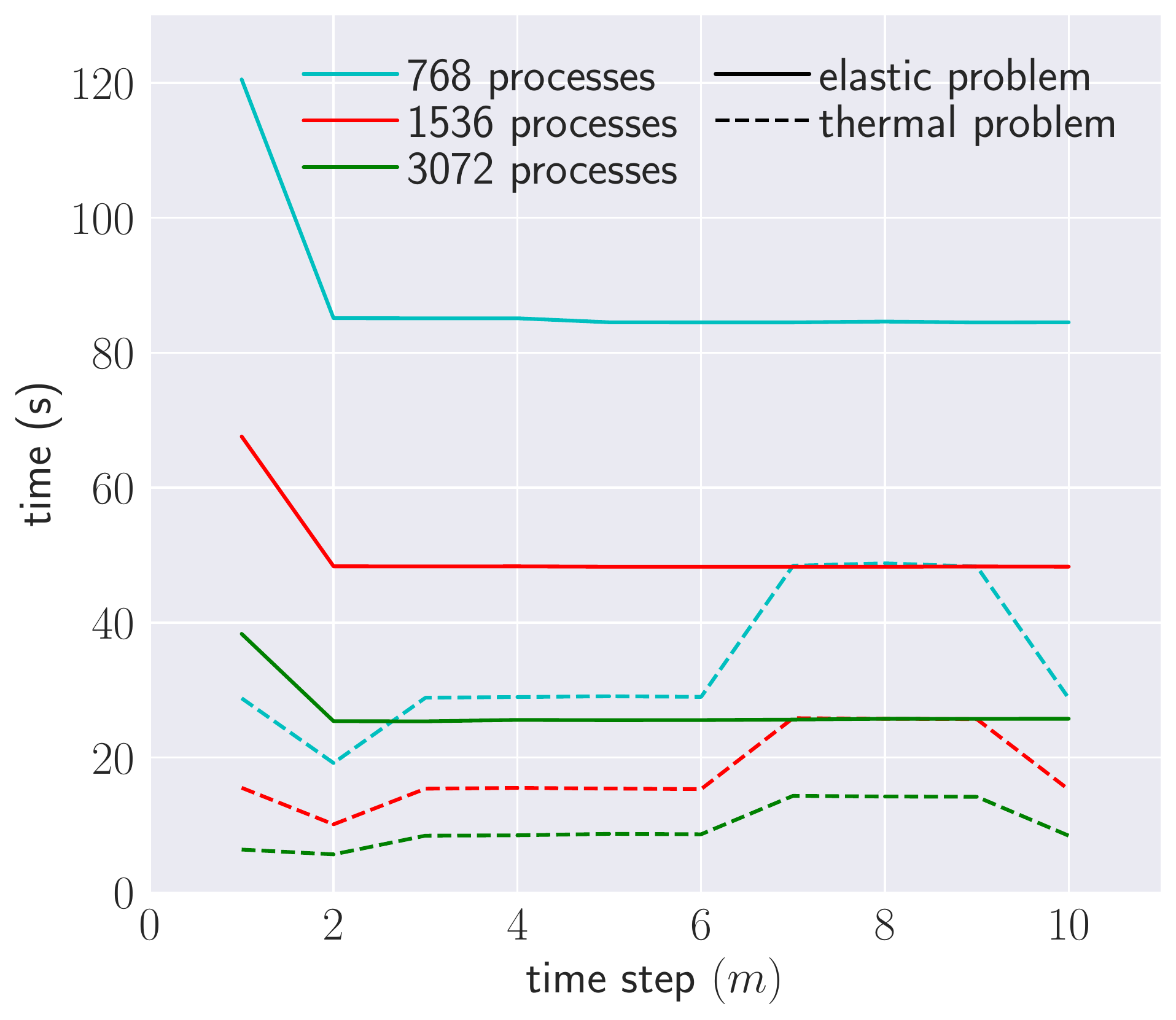}
    \caption{Computation time at each time step for differing numbers of processes.}
    \label{fig:tc_transient_strong_scaling_detail}
  \end{subfigure}%
  \begin{subfigure}[t]{0.49\textwidth}
    \includegraphics[height=0.25\textheight]{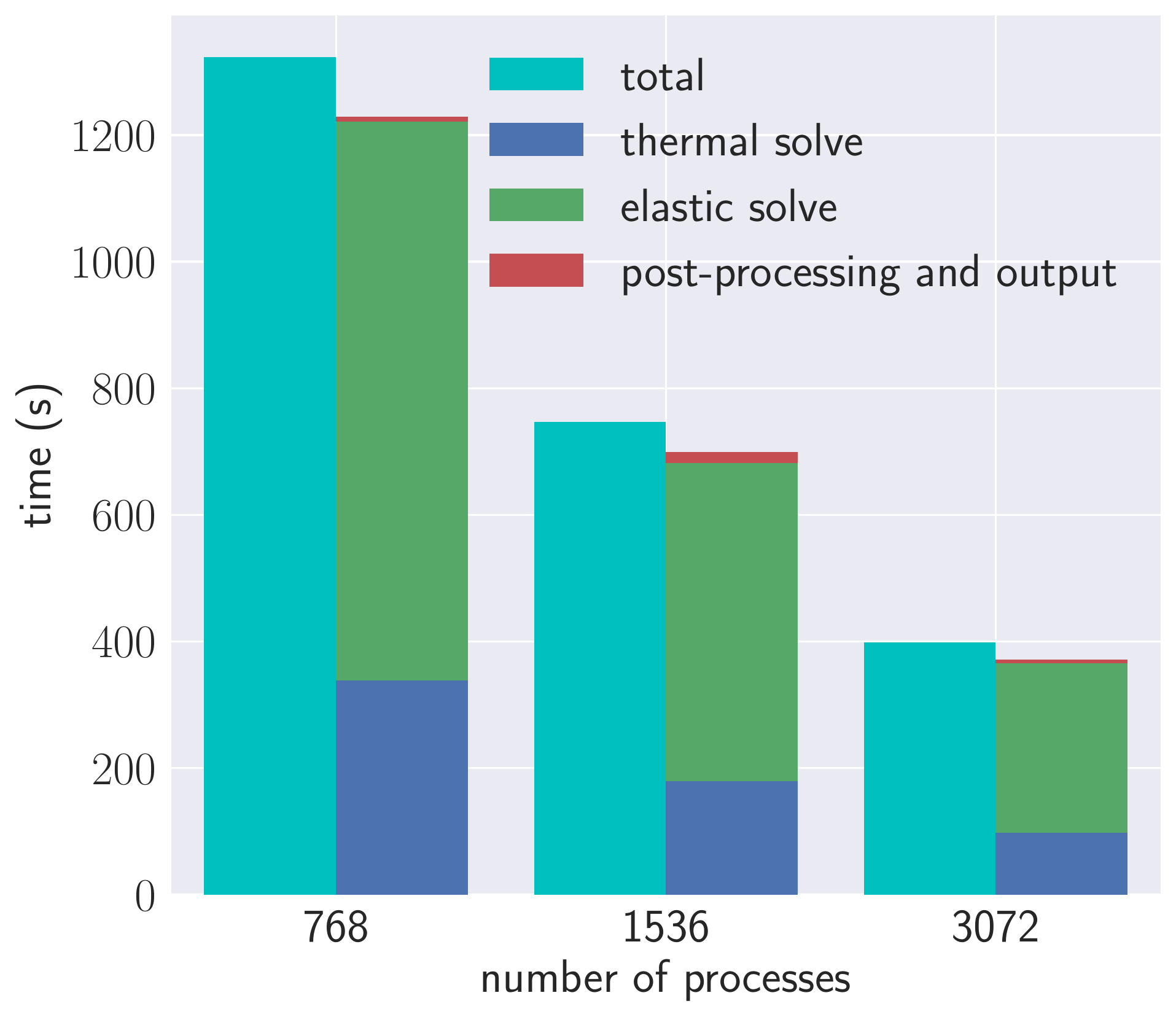}
    \caption{Cumulative computation time.}
    \label{fig:tc_transient_strong_scaling_cumulative}
  \end{subfigure}
  \caption{Strong scaling data for transient analysis of the
    turbocharger for \num{10} time steps using turbocharger mesh~`B' and
    $p = 2$. The problem has \num{67373812} thermal and \num{202121436}
    displacement degrees-of-freedom.}
  \label{fig:tc_transient_strong_scaling}
\end{figure}

Strong scaling for the transient turbocharger problems over 10 time
steps is presented in \cref{fig:tc_transient_strong_scaling}. We again
see good scaling behaviour, consistent with the steady analysis of the
turbocharger (cf.~\cref{fig:tc_steady_te_strong_scl_archer_p2}).
%------------------------------------------------------------------------------
\section{Conclusions}
\label{sec:conclusions}

We have demonstrated that it is possible to solve large-scale
thermomechanical turbomachinery problems scalably and efficiently using
iterative solvers. This is contrary to widely held views in the
turbomachinery community. Critical to the successful application of
iterative solvers are: (i) the selection of preconditioners that are
mathematically suitable for elliptic equations; (ii) proper
configuration of preconditioners using properties of the underlying
physical system, e.g. setting the near-nullspace for smoothed
aggregation AMG; and (iii) the use of high quality meshes. Iterative
solvers are less robust than direct solvers, and successful application
does require greater expertise and experience on the part of the
analyst, but they do offer the only avenue towards extreme scale
thermomechanical simulation.

The presented examples are representative of practical turbomachinery
simulations in terms of the materials and number of boundary conditions.
The results do show some areas where there is scope for performance
improvements, particularly for the solution of the mechanical problem in
terms of runtime and parallel scaling. It would be interesting to
investigate methods at higher process counts, and to explore methods
that have lower set-up cost and memory usage than smoothed aggregation
AMG. The presented examples are also simple, compared to the full range
of physical processes that are typically modelled in turbomachinery
analysis, such as geometrically nonlinear effects and contact. The
results in this work provide a platform that motivates research into
solver technology for a wider range of physical processes in the context
of thermomechanical simulation. The methods described and the tools used
are all available in open-source libraries. The implementations can be
freely inspected and used.

%------------------------------------------------------------------------------
\subsubsection*{Acknowledgements}

We thank Dr Mark Adams of Lawrence Berkeley National Laboratory for his
advice on the use of smoothed aggregation AMG. The support of Mitsubishi
Heavy Industries is gratefully acknowledged. CNR is supported by EPSRC
Grant EP/N018877/1.

%------------------------------------------------------------------------------
\bibliographystyle{abbrvnat}
\bibliography{references}
%------------------------------------------------------------------------------
\end{document}